\documentclass[aip, jmp, reprint, 11pt, onecolumn, showpacs, superscriptaddress, floatfix, nofootinbib, notitlepage]{revtex4-1}

\usepackage{amsthm,amsmath,amsfonts,amssymb,verbatim, color}
\usepackage{subfigure}
\usepackage{graphicx}
\usepackage[percent]{overpic}
\usepackage{bm}
\usepackage{epsfig,slashed}
\usepackage{epstopdf}
\usepackage{lipsum}
\usepackage{float}
\usepackage{anyfontsize}
\usepackage[colorlinks=true,citecolor=blue, linkcolor=blue,urlcolor=blue]{hyperref}
\usepackage{ragged2e}
\usepackage[margin=0.99in]{geometry}

\newcommand{\sgn}{\mathop{\mathrm{sgn}}}

\newcommand{\mb}{\bm}

\newcommand{\mc}{\mathcal}

\DeclareGraphicsExtensions{.png}

\begin{document}

\title{Exact eigenfunction amplitude distributions of integrable quantum billiards}
\author{Rhine Samajdar}
\affiliation{Department of Physics, Harvard University, Cambridge, MA 02138, USA}
\author{Sudhir R. Jain}
\email{srjain@barc.gov.in}
\affiliation{Nuclear Physics Division, Bhabha Atomic Research Centre, Mumbai 400085, India}
\affiliation{Homi Bhabha National Institute, Anushakti Nagar, Mumbai 400094, India}

\date{\today}

\begin{abstract}
\vspace*{0.1cm}
The exact probability distributions of the amplitudes of eigenfunctions, $\Psi(x, y)$, of several integrable planar billiards are analytically calculated and shown to possess singularities at $\Psi = 0$; the nature of this singularity is shape-dependent. In particular, we prove that the distribution function for a rectangular quantum billiard is proportional to the complete elliptic integral, ${\bf K}(1 - \Psi ^2)$, and demonstrate its universality, modulo a weak dependence on quantum numbers. On the other hand, we study the low-lying states of nonseparable, integrable triangular billiards and find the distributions thereof to be described by the Meijer G-function or certain hypergeometric functions. Our analysis captures a marked departure from the Gaussian distributions for chaotic billiards in its survey of the fluctuations of the eigenfunctions about $\Psi = 0$.
\end{abstract}
\maketitle

\section{Introduction}

In 1973, \citet{percival1973regular} introduced the classification of eigenfunctions, in the semiclassical limit, into regular and irregular according as whether the modes corresponded to integrable ray systems or classically nonintegrable Hamiltonians. Shortly thereafter, \citet{berry1977regular} suggested a simple quantitative method to distinguish between the two, proposing that the eigenfunctions of strongly chaotic systems behave like a random superposition of plane waves. This ansatz, based on eikonal theory, is motivated by the chaotic dynamics of the system as a result of which, a typical trajectory passes through an arbitrarily small neighborhood of every point. Mathematically, such a random superposition of plane waves on a region $\mathcal{D} \subset \mathbb{R}^2$ can be written as
\begin{equation}
\Psi_{\,\mathrm{RWM}}\, ({\bf x}) = \sqrt{\frac{2}{\mathrm{vol} (\mathcal{D})\, N}} \sum_{i = 1}^{N} a_i \, \cos\,({\bf k}_i \, {\bf x} + \phi_i),
\end{equation}
where $a_i \in \mathbb{R} $ are independent Gaussian random variables with zero mean and unit variance; $\phi_i$ are uniformly distributed random variables on $[0, 2 \pi)$, and the momenta $\mathbf{k}_i \in \mathbb{R}^2$ are randomly equidistributed, lying on a circle of radius $\sqrt{E}$.\\

The random wave model, in spite of lacking a rigorous proof, can be used to quantitatively estimate the statistical properties of wavefunctions in chaotic systems---perhaps the simplest instance of such a property would be the eigenfunction amplitude distribution. As has been argued by Refs.~\onlinecite{berry1977regular, berry1983semiclassical}, the local amplitude fluctuations of the eigenfunctions depend strongly on the underlying classical system. Using the central limit theorem, it is easily seen that in the random wave model, the probability of finding the value $\Psi$ at any point is distributed as a Gaussian, namely
\begin{equation}
P(\Psi) = \frac{1}{\sqrt{2\pi} \, \sigma}\,\exp \left( - \frac{\Psi^2}{2\, \sigma^2} \right),
\end{equation}
with $\sigma = 1 / \sqrt{\mathrm{vol} (\mathcal{D})}$ \cite{phdthesis}. Several numerical studies \cite{shapiro1984onset,mcdonald1988wave,o1988quantum,aurich1991exact,hejhal1992topography,aurich1993statistical,li1994statistical} have indeed confirmed the Gaussian amplitude distribution expected in a chaotic billiard.\\

The statistical properties of eigenfunctions have also gained particular importance in the context of disordered metals, especially in connection to localization effects. The distribution of eigenfunction amplitudes is known to be relevant to the description of fluctuations of tunneling conductance across quantum dots \cite{PhysRevLett.68.3468} and for understanding properties of atomic spectra \cite{PhysRevLett.70.1615}. Ignoring the spatial structure of the system, these statistics are well described, in the leading approximation, by random matrix theory (RMT), which again predicts a Gaussian distribution of amplitudes for systems with unbroken or completely-broken time-reversal symmetry \cite{mirlin1993statistics,efetov1993local}. The corrections to this distribution arising due to weak localization were evaluated by Ref.~\onlinecite{fyodorov1995mesoscopic} and shown to lead to ``\textit{an increase in probability to find a value of the amplitude considerably smaller or considerably larger than the average value}.'' The widespread interest in eigenfunction amplitude statistics highlighted by these observations emphasizes the significance of further theoretical investigation. Most of the efforts in this direction, however, have focused primarily on strongly chaotic systems. \citet{PhysRevE.63.046208} first explored a new intermediate regime of Wigner ergodicity (in which the states are nonergodic but the nearest-neighbor level-spacing statistics is still described by the Wigner distribution) in rough microwave billiards and found satisfactory agreement with the standard normalized Gaussian prediction, some deviations in the vicinity of zero notwithstanding. Soon, the random wave model was extended from the case of chaotic systems to systems with mixed phase space (where both regular and chaotic motions coexist \cite{markus1974generic}) by Ref.~\onlinecite{0305-4470-35-3-306}, which studied the amplitude distribution of irregular eigenfunctions in a lima{\c c}on billiard. The amplitude distribution for highly-excited, irregular eigenmodes of the $\pi /3$-rhombus  billiard (which do not vanish on either diagonal) too was found to be well approximated by a Gaussian \cite{piby3}. Conversely, the distribution functions for regular modes displayed a sharp rise near zero amplitude. These billiards are examples of nonchaotic, nonintegrable dynamical systems for which more detailed investigations are yet to be carried out.  \\ 

Regrettably, not much is known about the eigenfunction amplitude distributions of integrable systems and this article seeks to bridge the gap. What we do know already, however, is that, as \citet{berry1977regular} remarks, ``$\Psi$ \textit{for an integrable system cannot be a Gaussian random function because its spectrum of wavevectors is discrete}.'' Moreover, Ref.~\onlinecite{mcdonald1988wave} notes that a regular mode---which has underlying trajectories in phase space that are smoothly distributed on the scale of $\hbar$ or wavelength \cite{berry1977regular}---is characterized by a non-Gaussian probability distribution. The verity of this remark is also borne out for the regular modes of the $\pi /3$-rhombus billiard in the study mentioned above. These considerations hint that the amplitude distributions of integrable systems would, in general, be non-Gaussian and consequently, distinct from those predicted for chaotic ones. In the following sections, we present some of the first results in this regard for integrable systems. We compute explicit analytical expressions for the distribution functions of both separable and nonseparable integrable billiards in Secs.~\ref{Rect} and \ref{NS}, respectively. Our calculations illustrate the rich---and perhaps, surprising---mathematical structure of these new distributions, which belies the simplicity of the corresponding wavefunctions.

\section{Methodology}

In principle, our object of interest is well defined, namely, the probability density function (abbreviated as PDF hereafter) of $\Psi$. On the other hand, the approaches to this end documented in the literature are several and varied. For instance, the amplitude distribution can be evaluated in terms of correlation functions of a certain supermatrix $\sigma$ model \cite{efetov1983supersymmetry,verbaarschot1985grassmann,fyodorov1994statistical} such as in Ref.~\onlinecite{fyodorov1995mesoscopic}. Other studies, like Ref.~\onlinecite{mcdonald1988wave}, construct the probability distribution as a normalized histogram, with a specific number of bins, by sampling the normalized eigenfunctions at several points in the interior of $\mathcal{D}$. The choice of an appropriate method is thus somewhat arbitrary. However, in order to extract exact expressions for the PDF, it is advantageous to proceed by first computing the characteristic function (CF) \cite{feller2008introduction}. \\

\begin{figure}[htb]
\centering
\subfigure[]{
\label{fig:first}
\includegraphics[width=0.485\textwidth]{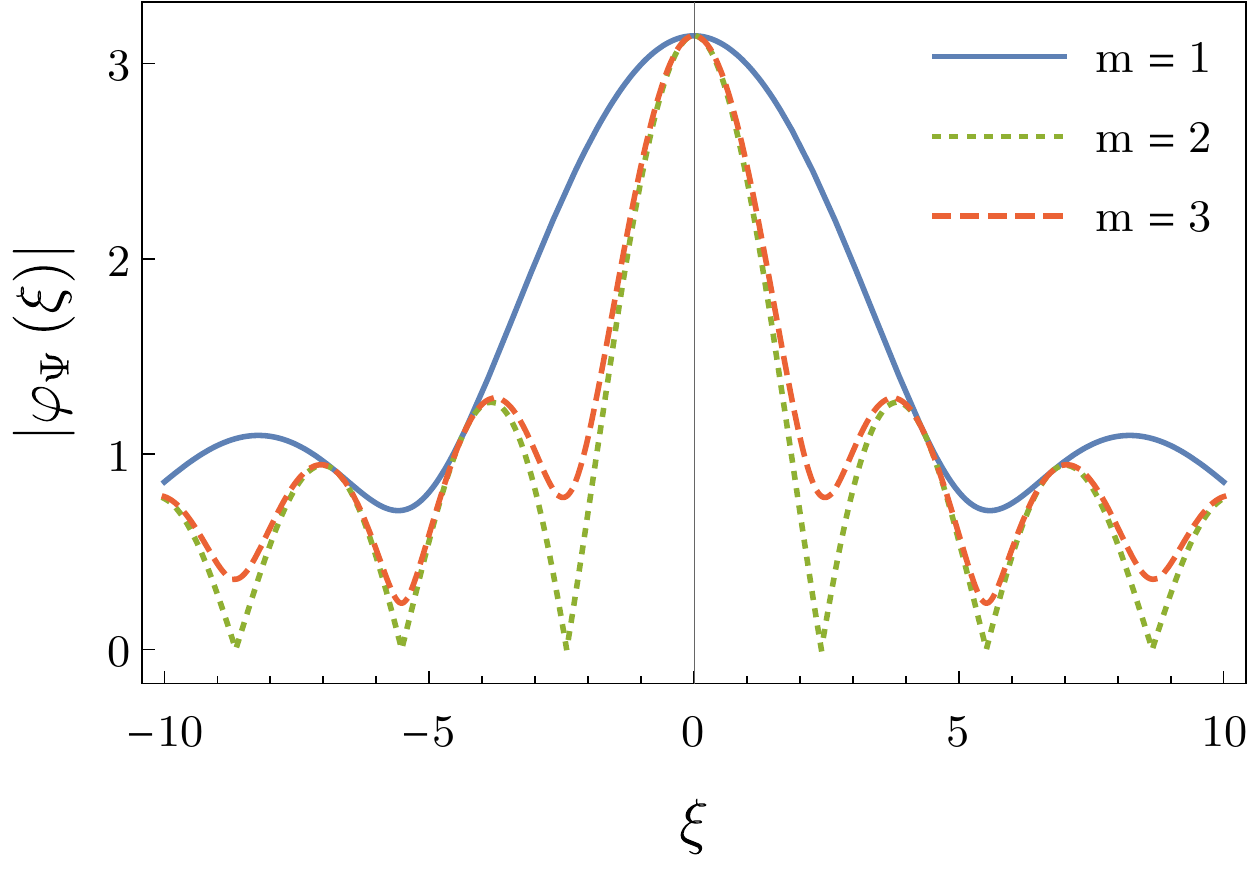}}
\subfigure[]{
\includegraphics[width=0.485\textwidth]{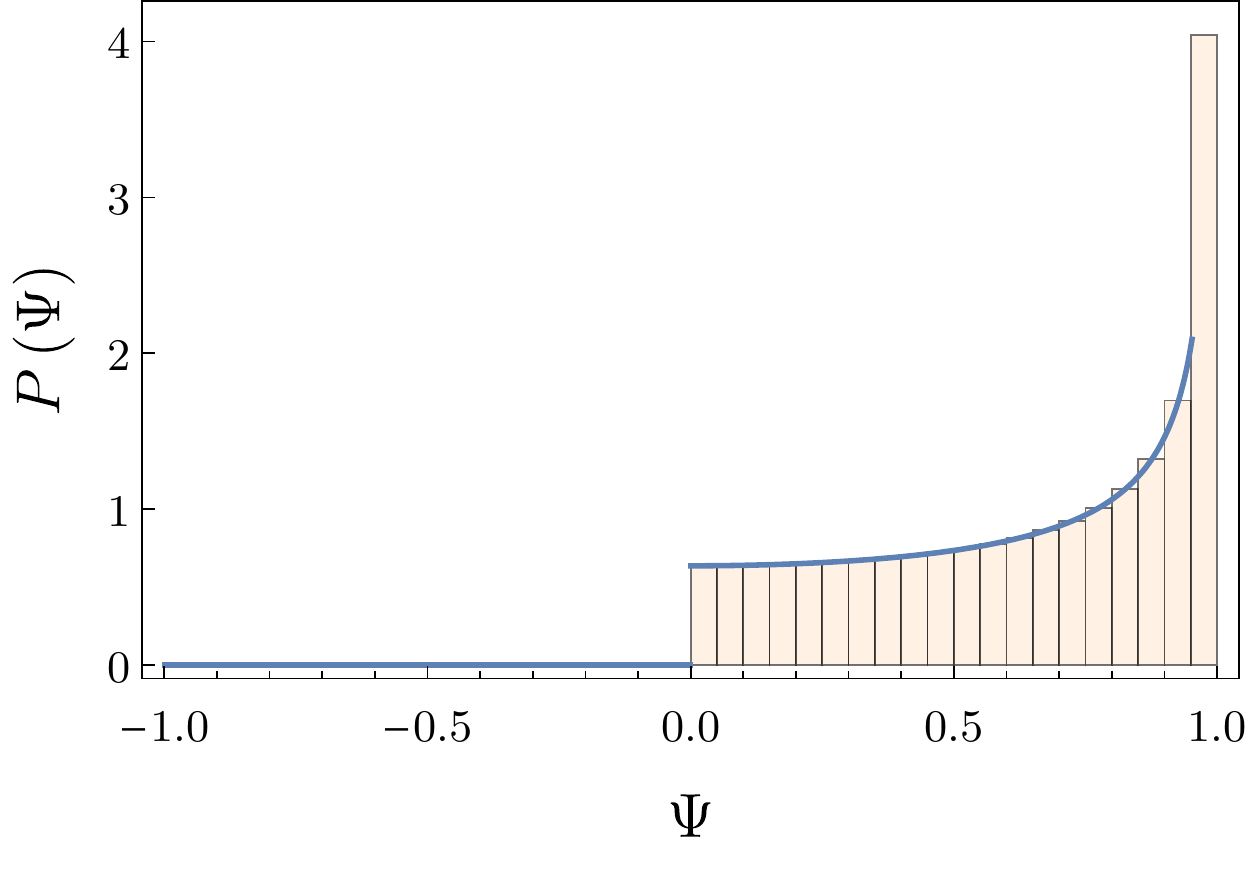}}
\subfigure[]{
\includegraphics[width=0.485\textwidth]{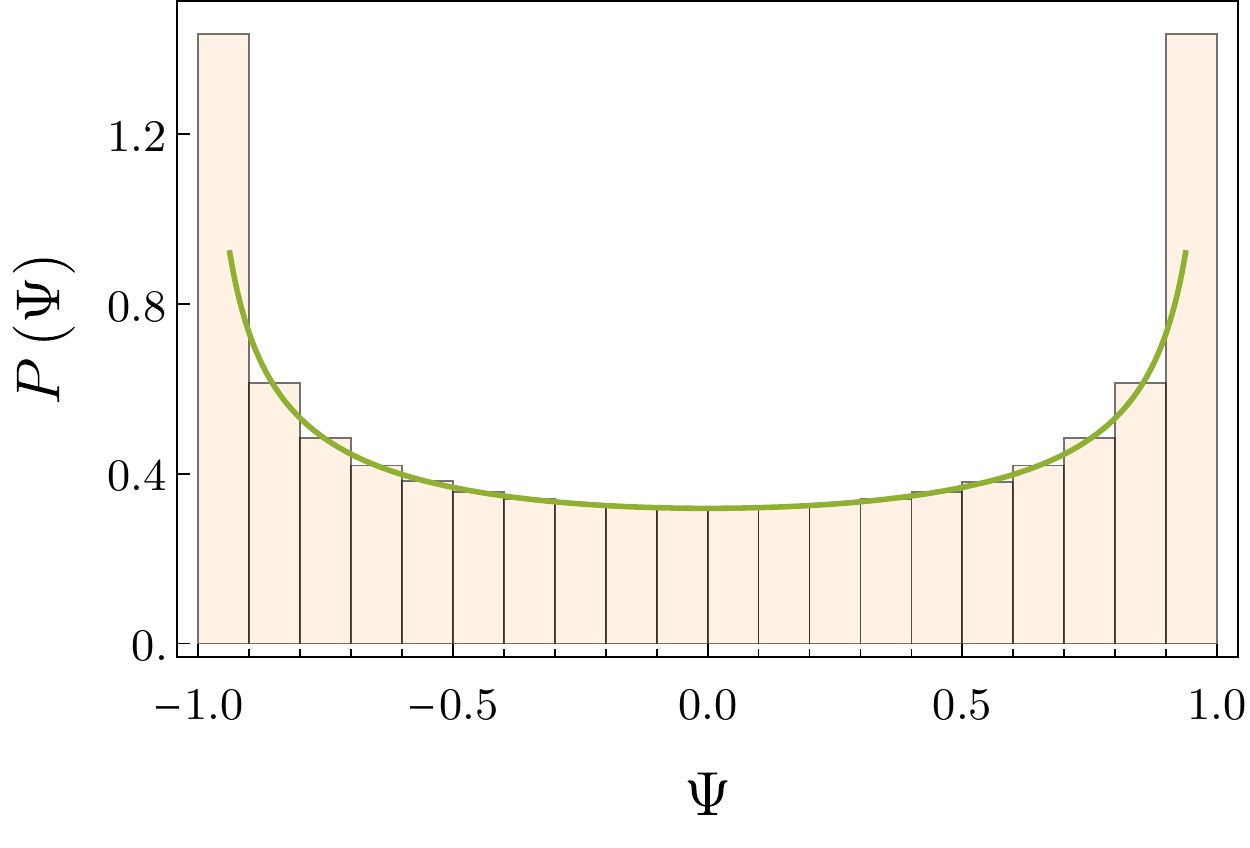}}
\subfigure[]{
\includegraphics[width=0.485\textwidth]{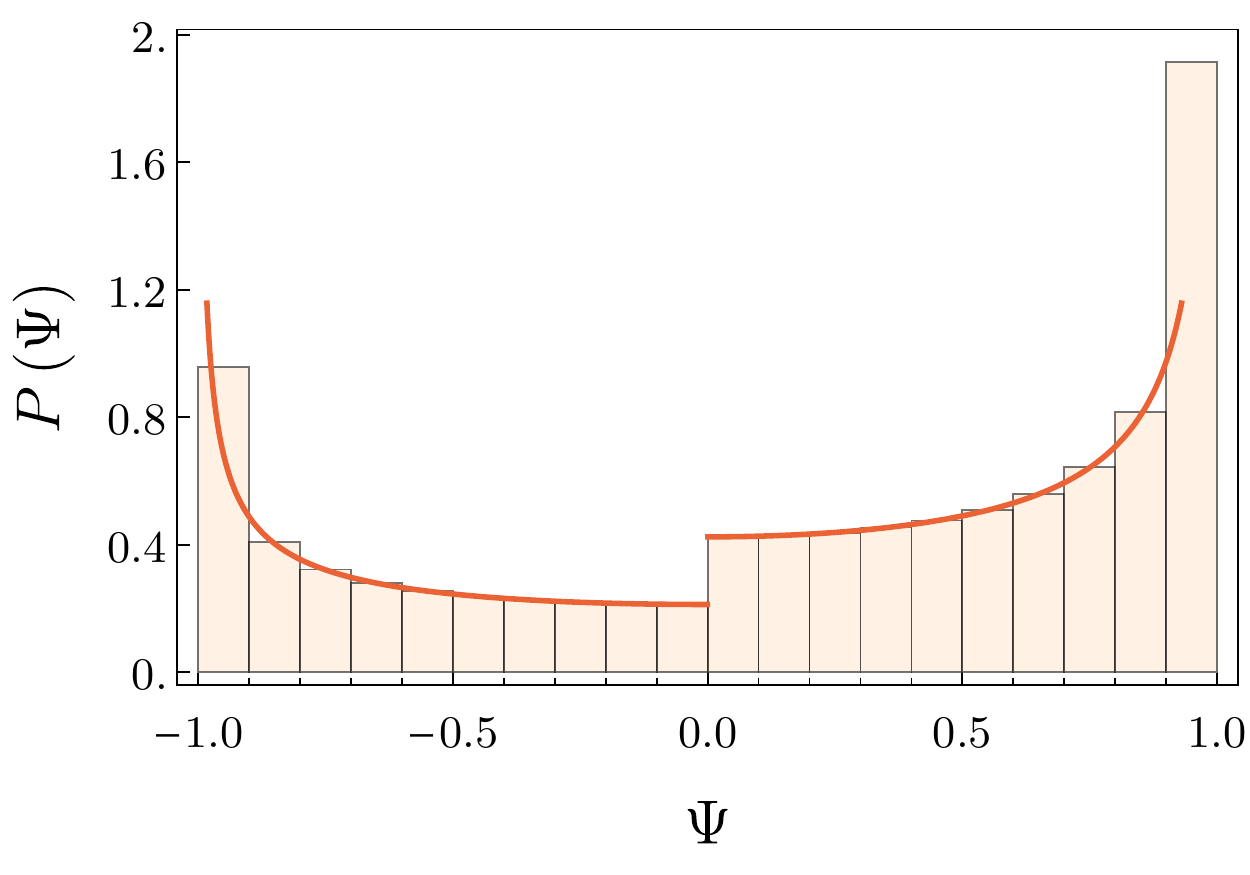}}
\caption{\label{fig:line}(a) The absolute value of the characteristic function (Eq.~\ref{eq:CF1D}) for the quantum numbers $m = 1$ to 3 for a particle in a one-dimensional well. The corresponding amplitude distributions of the wavefunctions (Eq.~\ref{eq:PDF1D}, solid lines) for $m=1$ (b), 2 (c), and 3 (d) demonstrate perfect agreement with the numerically-procured histograms.}
\end{figure}

To explicate this method, let us consider a particle on a line, confined to a one-dimensional ``hard-walled'' box $[0, \pi]$ with Dirichlet boundary conditions---the normalized wavefunctions of the system are simply $\Psi \, (x) =\sqrt{2/L} \, \sin\, (m\, x), \,\, m \in \mathbb{N}$. The CF is
\begin{equation}
\label{eq:CF1D}
\varphi_{\Psi} \, (\xi)\, = \int_{0 }^{\pi } \exp \big[ \,\mathrm{i}\, \xi  \,\sin\, (m x) \,\big]  \mathrm{d}x  = 
\begin{cases} 
\pi \, \bigg[ J_0\, (\xi) + {\displaystyle \frac{1}{m}} \,\mathrm{i} \,H_0\, (\xi) \bigg]; & \text{if m is odd,} \\[1em]
\pi \, J_0\, (\lvert \xi \rvert);       & \text{otherwise, }
\end{cases}
\end{equation}
where $J_\nu$ and $H_\nu$ stand for the Bessel function of the first kind and the Struve function, respectively, of order $\nu$. For the sake of simplicity, we temporarily ignore the normalization constant $\sqrt{2/L}$ in Eq.~\eqref{eq:CF1D} since the PDF of the normalized wavefunction can always be determined from that of the unnormalized one by scaling. The PDF of $\Psi$ is obtained from the Fourier transform of the CF, normalized by the length, as
\begin{equation}
\label{eq:PDF1D}
P \, (\Psi) = \frac{1}{\pi} \int_{-\infty }^{\infty }\,\exp\, ( -\mathrm{i}\, \xi  \,\Psi \,)\, \varphi_{\Psi} \, (\xi) \,\frac{\mathrm{d}\xi}{2\pi}  = 
\begin{cases} 
{\displaystyle \frac{m + \sgn\, (\Psi)}{m\, \pi \, \sqrt{1 - \Psi^2}}}; & \text{if m is odd,} \\[1em]
{\displaystyle \frac{1}{\pi \, \sqrt{1 - \Psi^2}}};      & \text{otherwise. }
\end{cases}
\end{equation}
As Fig.~\ref{fig:line} shows, this distribution exhibits a minimum at $\Psi = 0$ and diverges at $\Psi = \pm1 $, in sharp contrast to the behavior of the PDF for billiards in two dimensions calculated in the following sections.

\section{Separable systems}
\label{Rect}

The two-dimensional Helmholtz equation is separable in exactly four coordinate systems: the Cartesian, polar, elliptic, and parabolic coordinates \cite{eisenhart1934separable,eisenhart1948enumeration}. Separability, as the name suggests, implies that the wavefunction of the billiard can be obtained by separation of variables, and consequently, expressed as a  product of functions of independent variables. The prototypical separable system is the rectangular billiard $\mathcal{D} = [0, \pi] \times [0, \pi]$,\footnote{It is not difficult to see that the final distribution function (Eq.~\ref{eq:PDFRect}) holds for \textsl{any} rectangular box with arbitrary dimensions, as is evident from a trivial rescaling of variables: $x \rightarrow X = x\, \pi / L_x$, $y \rightarrow Y = y\, \pi /L_y$.} the eigenfunctions of which, with Dirichlet boundary conditions along the periphery $\partial \mathcal{D}$, are standing waves given by $\Psi \, (x,y) =(2/\pi) \, \sin\, (m\, x) \, \sin\, (n\, y), \,\, m, n \in \mathbb{N}$. The associated energy eigenvalues are $E_{m,n} \sim m^2 + n^2$. As reasoned previously, we may, without loss of generality, neglect the normalization factor in calculating the CF $\varphi_{\Psi} \, (\xi)$; it is therefore subsequently implicit that $P \, (\Psi) = 0$ for $\lvert \Psi \rvert > 1$. Akin to Eq.~\eqref{eq:CF1D}, there are two cases depending on the parity of the quantum numbers; we examine both individually. When $m$ is odd,
\begin{eqnarray}
\nonumber \varphi_{\Psi} \, (\xi)\, &=& \int_{0 }^{\pi }\int_{0 }^{\pi } \exp \big[ \,\mathrm{i}\, \xi  \,\sin\, (m\, x) \, \sin\, (n \,y) \,\big] \, \mathrm{d}x\,\mathrm{d}y \\
\label{eq:CFRect}
\label{eq:CFRect2}
&=& \int_{0 }^{\pi/2n } 2\pi \, \bigg[ n\,J_0\,  \left (\xi\,\sin\, (n\, y) \right) + {\displaystyle \frac{1}{m}} \,\mathrm{i} \,H_0\,  \left (\xi\,\sin\, (n\, y) \right) \, \delta_{n\, (\mathrm{mod}\, 2), 1} \bigg] \,\mathrm{d}y\\
\nonumber&=& \int_{0 }^{1} \frac{2\pi}{n\,\sqrt{1-u^2}} \bigg[ n\, J_0\, (\xi\,u) + {\displaystyle \frac{1}{m}} \,\mathrm{i} \,H_0\, (\xi\,u) \, \delta_{n\, (\mathrm{mod}\, 2), 1} \bigg] \mathrm{d}u  \equiv \mc{I}_1\, (\xi) + \mathrm{i}\, \mc{I}_2 \,(\xi) \, \delta_{n\, (\mathrm{mod}\, 2), 1},
\end{eqnarray}
wherein we have employed a change of variables to $u = \sin \,(n\,y)$. The Kronecker delta in Eq.~\eqref{eq:CFRect2} necessitates $n$ being odd for a nonvanishing contribution to $\varphi_{\Psi} \, (\xi)$ from the Struve function. \\

The two pieces of the integral in Eq.~\eqref{eq:CFRect2} are best evaluated using the power-series representations of the Bessel \cite{olver1965bessel} and Struve functions \cite{olver2010nist}:
\begin{alignat}{2}
\nonumber \mc{I}_1\, (\xi) &= 2\pi \int_{0 }^{1} \sum_{t \,= \,0}^\infty  \frac{(-1)^t}{t!\,  \Gamma (t+1)} \bigg(\frac{1}{4}\, \xi^2 u^2 \bigg)^t \frac{\mathrm{d}u}{\sqrt{1-u^2}}
&&= 2\pi\hspace*{-0.25cm}\sum_{t = 0,2,4,\ldots}^{\infty} {\displaystyle \frac{(-1)^{t/2}}{\frac{t}{2}!\, \Gamma \big(\frac{t}{2}+1\big)}}\int_{0 }^{1}\bigg(\frac{1}{2}\, \xi\, u \bigg)^t \frac{\mathrm{d}u}{\sqrt{1-u^2}}\\
&= \pi^{3/2} \sum_{t \,= \,0,2,4,\ldots}^{\infty}  \frac{(-1)^{t/2}}{(t/2)!}\, \, \frac{\Gamma \big(\frac{t+1}{2}\big)}{\Gamma \big(\frac{t}{2}+1\big)^2} \,\bigg(\frac{1}{2}\, \xi\, \bigg)^t &&= \pi^{2} \, J_0\, \bigg(\frac{\xi}{2} \bigg)^2.
\end{alignat}
The Fourier transform of this function can be computed using readily-available tables of integrals of double products of Bessel functions \cite{gradshteyn2014table}, thereby obtaining
\begin{equation}
\frac{1}{2\pi}\int_{-\infty }^{\infty }  \,\exp\, ( -\mathrm{i}\, \xi  \,\Psi \,)\,\,\mc{I}_1 \, (\xi) \, \mathrm{d}\xi = 2\,  \mb{K}\left(1-\Psi ^2\right).
\end{equation}
Here, $\mb{K}$ represents the complete elliptic integral of the first kind (which can be determined to arbitrary numerical precision) and the values thereof are well known in the literature \cite{hammersley1953tables}. Similarly,
\begin{alignat}{2}
\nonumber \mc{I}_2 \, (\xi)&= \frac{4}{m\, n} \int_{0 }^{1} \sum_{t \,= \,1,3,5,\ldots}\,  \frac{(-1)^{(t-1)/2}}{1^2\cdot3^2 \cdots t^2} \,(\xi\, u)^t \frac{\mathrm{d}u}{\sqrt{1-u^2}}\\
&= \frac{2 \sqrt{\pi}}{m\, n} \sum_{t \,= \,1,3,5,\ldots}\,  \frac{(-1)^{(t-1)/2}}{1^2\cdot3^2 \cdots t^2} \,\, \frac{\Gamma \big(\frac{t+1}{2}\big)}{\Gamma \big(\frac{t}{2}+1\big)^2}\,\, \xi^t
=  \frac{4}{m\, n}\, \xi \, \, _2F_3\left(1,1; \,\frac{3}{2},\frac{3}{2},\frac{3}{2}; \,-\frac{\xi ^2}{4}\right),
\end{alignat}
$F$ being the generalized hypergeometric function with the series expansion 
\begin{equation*}
_qF _p (a;b;z) = \Sigma _{k=0}^{\infty } (a_1)_k \ldots (a_p)_k / (b_1)_k \ldots (b_q)_k\, z^k/ k!,
\end{equation*}
expressed in terms of the Pochhammer symbol $(a)_k$. The corresponding Fourier transform is \cite{gradshteyn2014table}
\begin{equation}
\frac{1}{2\pi}\int_{-\infty }^{\infty }  \,\exp\, ( -\mathrm{i}\, \xi  \,\Psi \,)\,\,\mc{I}_2 \, (\xi) \, \mathrm{d}\xi = -\mathrm{i}\,\frac{2}{m\,n}\,\frac{ \mb{K} \left(1-\Psi ^2\right)}{\text{sgn}\,(\Psi )}.
\end{equation}
An exactly analogous approach can be adopted for the situation where $m$ is even:
\begin{eqnarray}
\label{eq:CFRect3}
\nonumber \varphi_{\Psi} \, (\xi)\, &=& \int_{0 }^{\pi } \pi \, J_0\, (\lvert \, \xi\,\sin n y \rvert) \,\mathrm{d}y 
= \int_{0 }^{\pi/2n } 2\pi \, n\,J_0\, (\lvert \, \xi\,\sin n y \rvert) \,\mathrm{d}y
= \int_{0 }^{1} 2\pi\,\frac{\, J_0\, ( \lvert \xi \rvert \,u)}{\sqrt{1-u^2}} \,\mathrm{d}u\\
&=& \pi^{2} \, J_0\, \big(\lvert \xi \rvert/2 \big)^2,
\end{eqnarray}
and its Fourier transform is, once again, $ 2\,  \mb{K}\left(1-\Psi ^2\right)$. Reassembling all the components, we summarize the final distribution function,\footnote{It has been brought to our notice that a similar result was obtained independently by \citet{beugeling2017statistical} after the submission of the present manuscript.} normalized by the area of the square billiard:
\begin{equation}
P \, (\Psi) = \frac{1}{\pi^2} \int_{-\infty }^{\infty } \,\exp\, ( -\mathrm{i}\, \xi  \,\Psi \,)\, \varphi_{\Psi} \, (\xi) \, \frac{\mathrm{d}\xi}{2\pi} = 
\begin{cases} 
{\displaystyle \frac{2}{\pi^2}\,  \mb{K}\left(1-\Psi ^2\right) \bigg[1 + \frac{\sgn\, {\Psi}}{m\,n} \bigg]}; & \text{$m$, $n$ are odd,} \\[1em]
{\displaystyle \frac{2}{\pi^2}\,  \mb{K}\left(1-\Psi ^2\right)} ;  & \text{otherwise},
\label{eq:PDFRect}
\end{cases}
\end{equation}

\begin{figure}[htb]
\centering
\subfigure[]{
\includegraphics[width=0.484\textwidth]{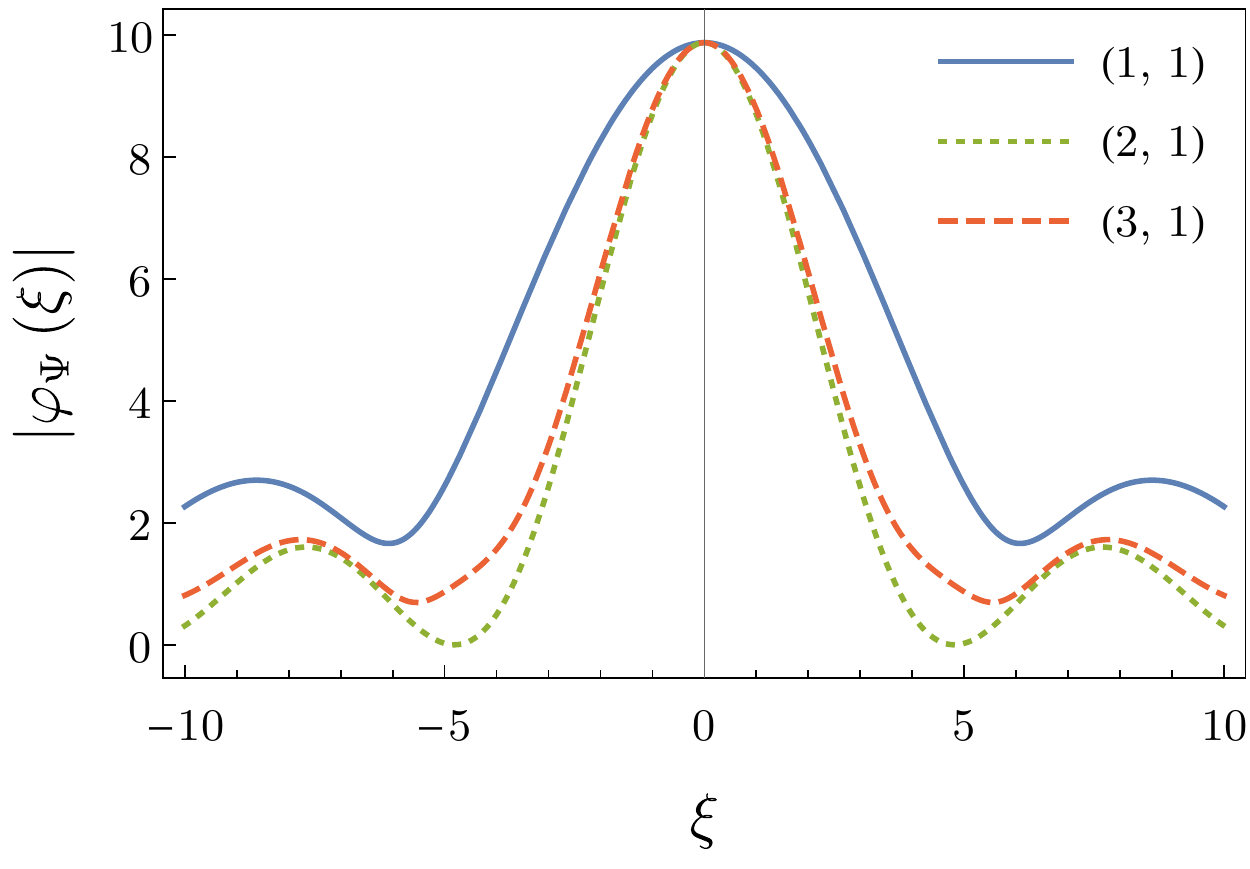}}
\subfigure[]{
\includegraphics[width=0.491\textwidth, trim={0 0 0 0.01cm}, clip]{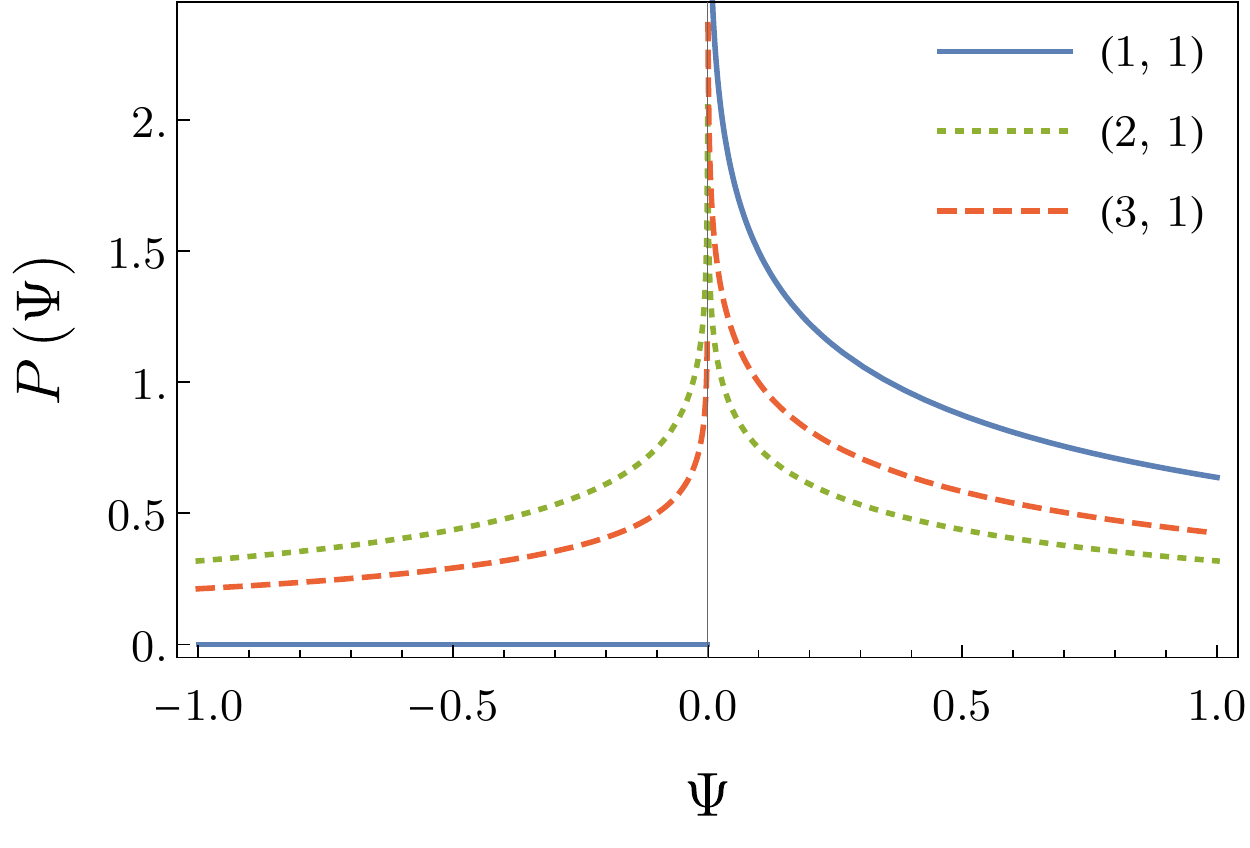}}
\caption{(a) The absolute value of the characteristic function, and (b) the probability distribution function of the wavefunction amplitudes for different states of the square billiard, labelled by $(m, n)$. Note that the curves marked $(2,1)$ are identical to those for any eigenstate with at least one even quantum number.}
\end{figure}

\begin{figure}[htb]
\centering
\subfigure[]{
\label{fig:Rect1}
\includegraphics[width=0.25\textwidth]{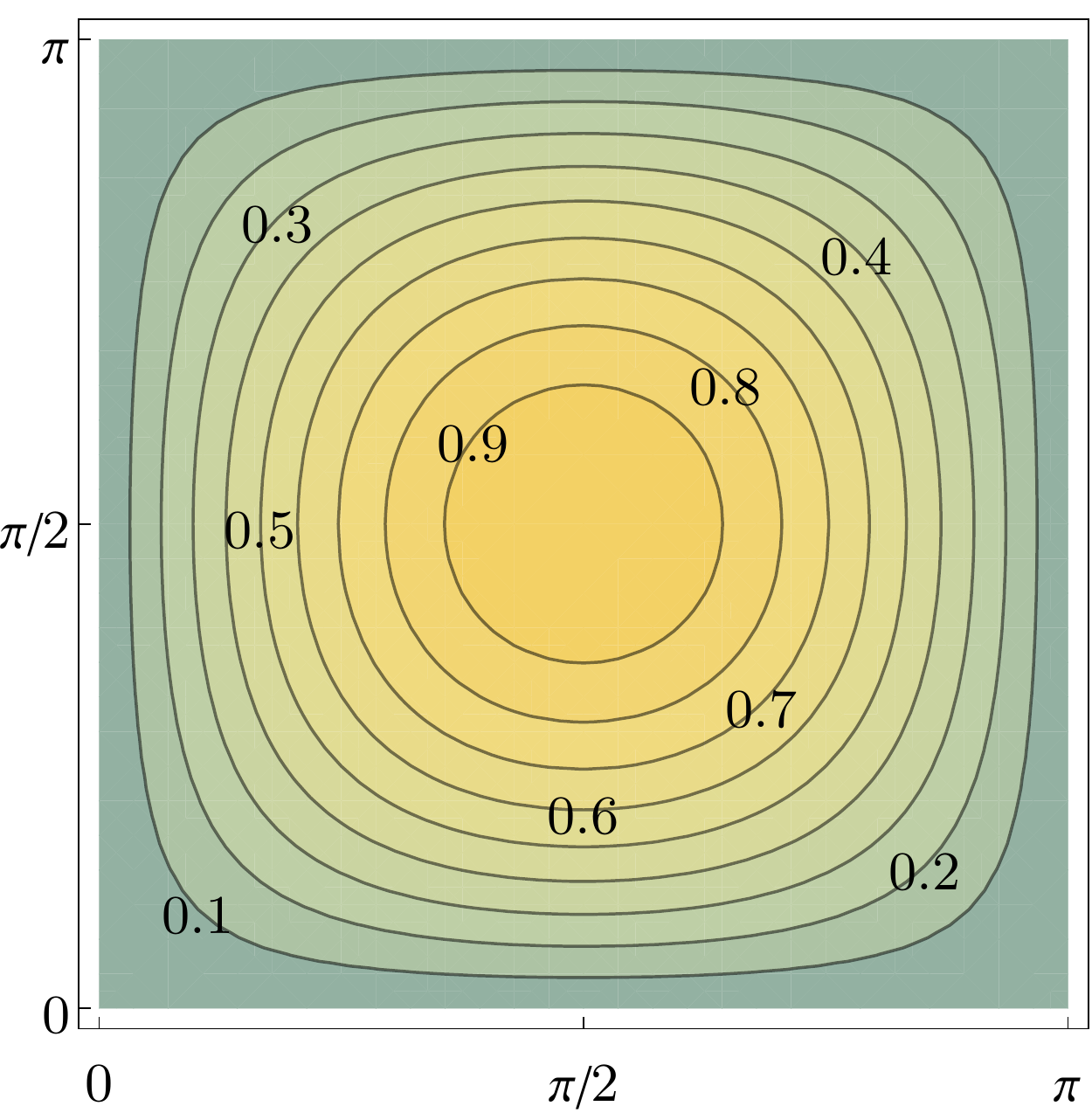}}\quad\quad
\subfigure[]{
\includegraphics[width=0.25\textwidth]{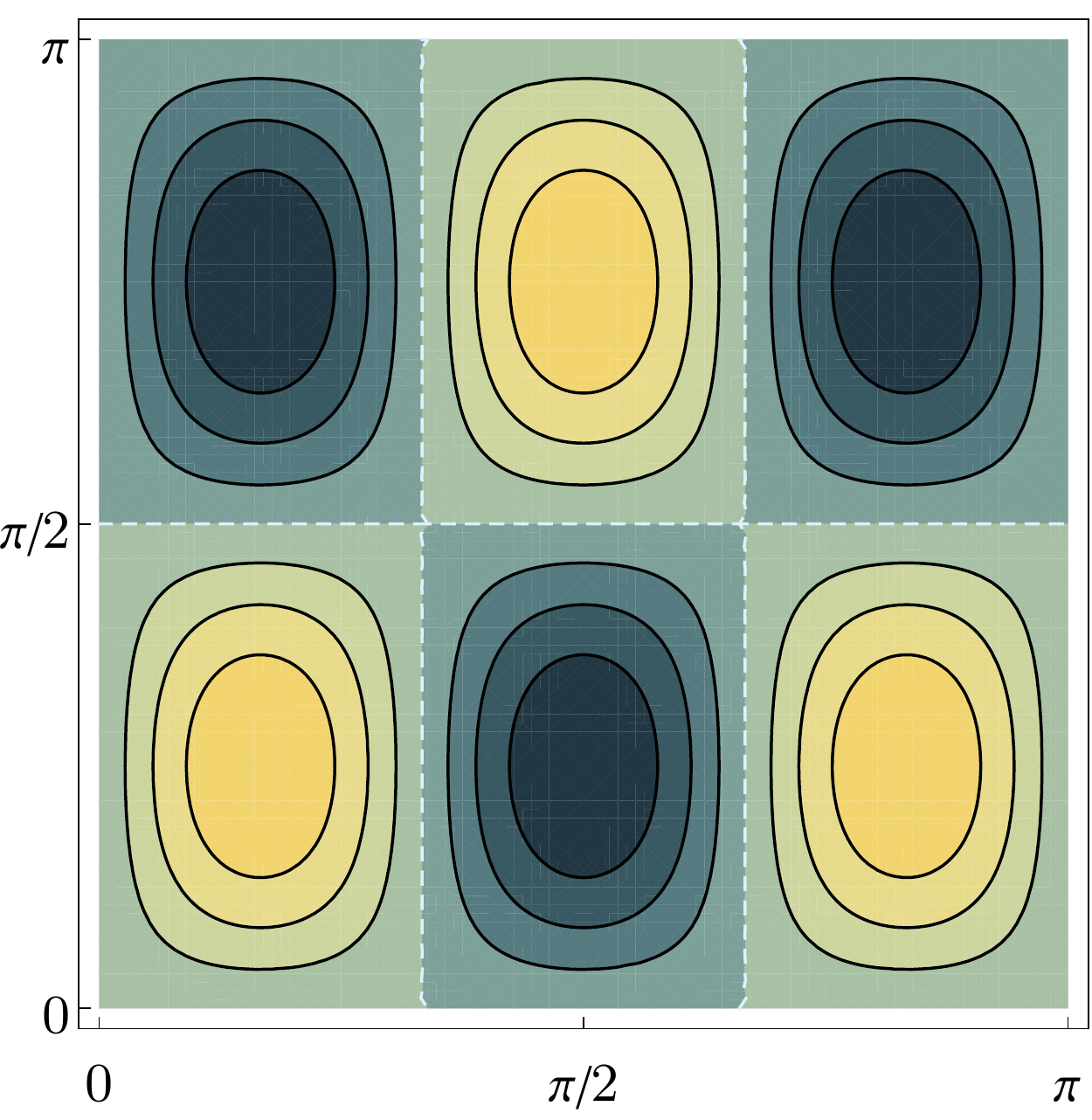}}\quad\quad
\subfigure[]{
\includegraphics[width=0.25\textwidth]{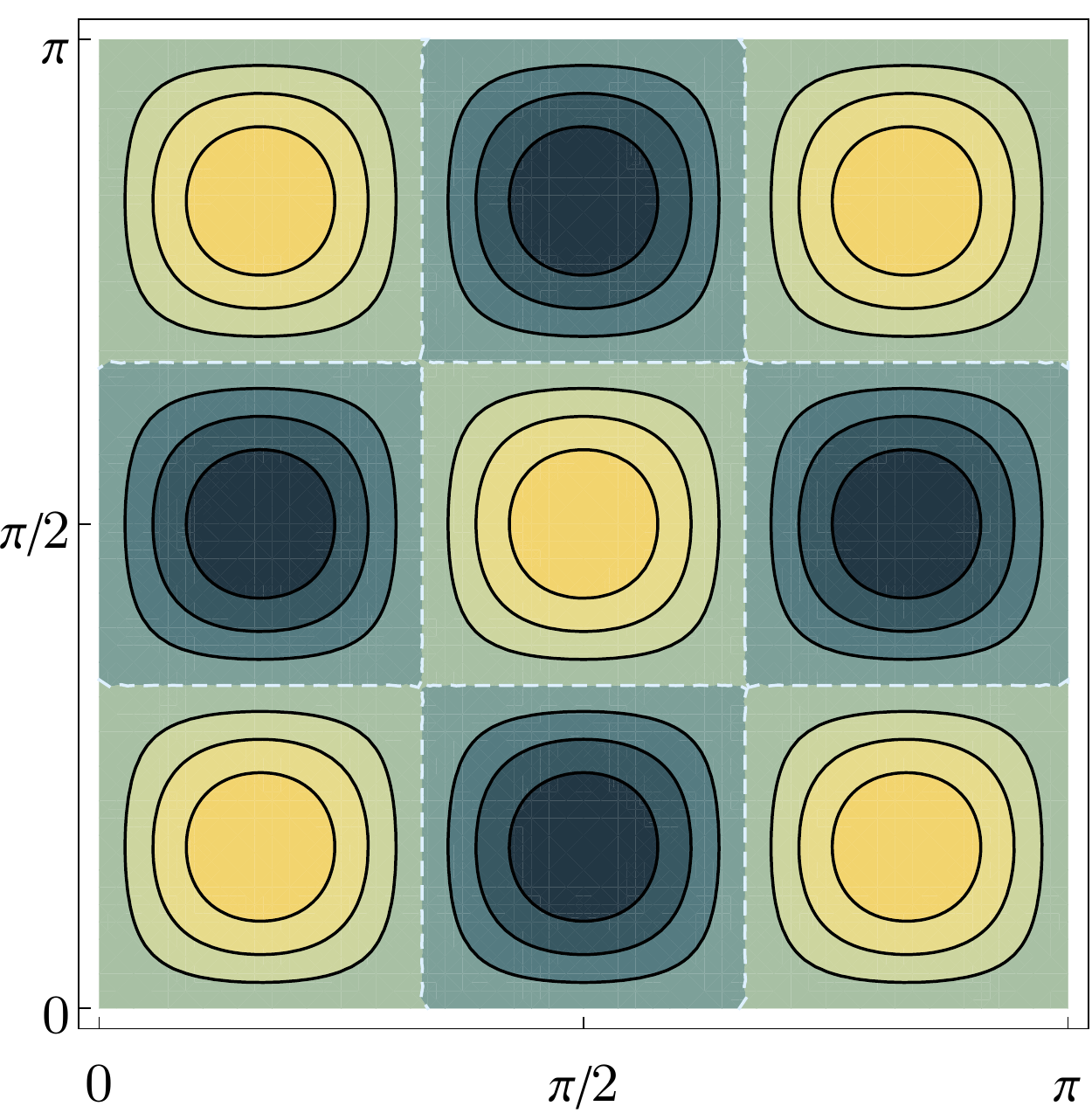}}\quad\quad
\subfigure{
\includegraphics[height=1.55in]{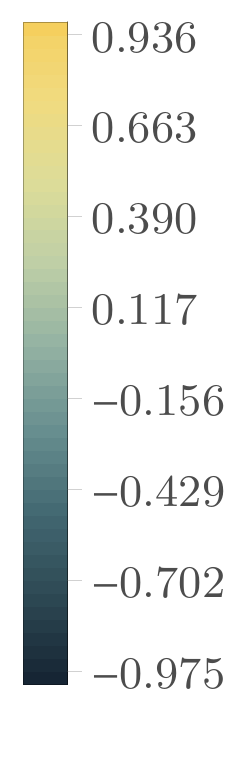}}
\caption{\label{fig:rect}Contour plots of the wavefunctions of a square quantum billiard; the nodal domains are demarcated by dashed white lines. The bright (dark) patches correspond to regions where the wavefunction is positive (negative). (a) The eigenfunction of the ground state, $(1,1)$, is entirely positive. (b) With one even quantum number, the state $(3,2)$ has an equal number of equiareal positive and negative domains. (c) The state $(3,3)$ (with both $m$ and $n$ odd) has one residual (positive) domain, which asymmetrically skews the PDF.}
\end{figure}

A feature of the PDF that merits elaboration is the absence of any dependence on the quantum numbers $m$ and $n$, unless they both happen to be odd. Moreover, the distribution, barring the first case of Eq.~\eqref{eq:PDFRect} above, is symmetric under the transformation $\Psi \rightarrow -\Psi$ as well. The asymmetry in the distribution when both $m$ and $n$ are odd can be accounted for by examining the pattern of the nodal domains \cite{jain2017nodal}---the maximally connected regions on the manifold $\mc{D}$ where the wavefunction does not change sign. The wavefunction of the rectangle, as can be observed in Fig.~\ref{fig:rect}, consists of a grid of intersecting nodal lines forming a ``checkerboard'' pattern. The number of such nodal domains for a rectangular billiard is just the product of the two quantum numbers. If either $m$ or $n$ is even, so is the number of domains and there are an equal number of positive and negative domains---this enforces the symmetry of $P\, (\Psi)$ about $\Psi = 0$. However, when the product $m \, n$ is an odd integer, there exists one additional positive/negative domain, without any counterpart of the opposite sign, and consequently, the symmetry of the eigenfunction amplitudes is broken. It is interesting to note that this symmetry breaking is essential to ensure that $P\, (\Psi) = 0\,\, \forall\,\, \Psi < 0$ in the ground state $m = n = 1$ (in which the wavefunction is positive throughout; see Fig.~\ref{fig:Rect1}). Nevertheless, in the limit of large quantum numbers, the presence of $m\,n \, (\sim E$) in the denominator rapidly eliminates the antisymmetric term and the distribution is asymptotically invariant under $\Psi \rightarrow -\Psi$. The divergence as $\Psi \rightarrow 0$, however, persists in the semiclassical limit---this statement also holds for other separable, integrable systems such as the circular billiard, as sketched in Appendix~\ref{app}.

\section{Nonseparable billiards}
\label{NS}

The only integrable but nonseparable billiards in two dimensions are the right-angled isosceles, the equilateral, and the $30^{\circ}-60^{\circ}-90^{\circ}$ hemiequilateral triangle \cite{Schachner1994,kaufman1999}. The wavefunctions of these billiards often assume remarkably complicated forms with labyrinthine networks of nodal lines and avoided crossings \cite{uhlenbeck1976generic}, quite unlike the separable systems discussed up till now. A consequence of this complexity is that it now becomes nigh impossible to analytically calculate the CF or PDF for any wavefunction beyond the ground or first-excited states. Even though we concentrate on primarily these low-lying states in this section, our analysis enables us to still glean some of the more general features of the distributions manifested by such systems. 

\subsection{Right-angled isosceles triangle}
The (unnormalized) wavefunctions for a right-isosceles triangle with each equal side of length $\pi$, defined by the region ${\cal D} = \big \{ (x, y) \in [0, \pi]^2: y \le x \big\}$, are 
\begin{equation}
\Psi_{m, n} (x, y) = \sin\, (m\,x)\, \sin\, (n\,y) - \sin\, (n\,x)\, \sin\, (m\,y),
\label{eq:Iso}
\end{equation}
where $m, n \in \mathbb{N}$ are two integer quantum numbers that determine the spectrum; the corresponding eigenvalues are $E_{m,n} = m^2 + n^2$, incidentally, no different from those for the separable square.

\subsubsection{Ground state}

The ground state wavefunction corresponds to the quantum numbers $m = 1$ and $n = 2$ (or vice versa). As usual, the CF is
\begin{equation}
\label{eq:IscCF1}
\varphi_{\Psi} \, (\xi)\, = \int_{0 }^{\pi }\int_{0 }^{x} \exp \big[ \,\mathrm{i}\, \xi  \, \left( \sin\, x \, \sin\, 2 y - \sin\, 2 x \, \sin\, y \right) \,\big] \, \mathrm{d}y\,\mathrm{d}x. 
\end{equation}
From a practical perspective, it is rather inconvenient to integrate over the triangular domain owing to the $x$-dependence of the limits thereof. Hence, it is easiest to first extend the region of integration from $[0, \pi] \times [0, x]$ to the square $[0, \pi] \times [0, \pi]$ and subsequently, eliminate the spurious contributions. To this end, we rewrite Eq.~\eqref{eq:IscCF1} as
\begin{alignat}{3}
\label{eq:CFSplit1}
\varphi_{\Psi} \, (\xi) &= \frac{1}{2}&&\sum_{t=0, 2, 4, \ldots}^{\infty} \frac{(\mathrm{i}\,\xi)^{t}}{t!}\,\int_{0 }^{\pi }\int_{0 }^{\pi} \left( \sin\, x \, \sin\, 2 y - \sin\, 2 x \, \sin\, y \right)^{t} \, \mathrm{d}x\,\mathrm{d}y \quad &&(\equiv \mathcal{I}_1 \,(\xi))\\
\label{eq:CFSplit2}
&+ &&\sum_{t=1, 3, 5, \ldots}^{\infty} \frac{(\mathrm{i}\,\xi)^{t}}{t!}\,\int_{0 }^{\pi }\int_{0 }^{x} \left( \sin\, x \, \sin\, 2 y - \sin\, 2 x \, \sin\, y \right)^{t} \, \mathrm{d}y\,\mathrm{d}x \quad &&(\equiv \mathcal{I}_2 \,(\xi)).
\end{alignat}
Thus restructured, Eq.~\eqref{eq:IscCF1} stands more amenable to manipulation. Now,
\begin{alignat*}{1}
\mathcal{I}_1\,(\xi) &= \sum_{t=0, 2, 4, \ldots}^{\infty} \frac{(-1)^{t/2}\,\xi^t}{2\,\,t !}\,\int_{0 }^{\pi }\int_{0 }^{\pi} \sum_{k = 0}^{t} (-1)^k\,  \binom{t}{k} \sin^k x \, \sin^k 2 y\, \sin^{t-k} 2 x \, \sin^{t-k} y \,\, \mathrm{d}x\,\mathrm{d}y\\
&= \sum_{t=0, 2, 4, \ldots}^{\infty} \frac{(-1)^{t/2}\, 2^{t+1}\,\xi^t}{t !}\sum_{k = 0, 2, 4, \ldots}^{t}\,\int_{0 }^{\pi/2}\int_{0 }^{\pi/2} \binom{t}{k} \sin^t x \, \cos^{t-k} x\, \sin^t y \, \cos^k y \,\, \mathrm{d}x\,\mathrm{d}y\\
&= \sum_{t=0, 2, 4, \ldots}^{\infty} \frac{(-1)^{t/2}\, 2^{t+1}\,\xi^t}{t !}\sum_{k = 0, 2, 4, \ldots}^{t} \binom{t}{k}\, \frac{1}{2}\textrm{B} \bigg( \frac{t+1}{2}, \frac{t-k+1}{2} \bigg) \cdot \frac{1}{2}\textrm{B} \bigg(\frac{t+1}{2}, \frac{k+1}{2} \bigg),
\end{alignat*}
where $\textrm{B} (x, y)$ is the beta function (Euler's integral of the first kind). Identifying $t = 2 s$ and $k = 2j$, we have
\begin{alignat}{1}
\nonumber &\sum_{s=0}^{\infty} \frac{(-1)^{s}\, 2^{2s-1}\,\xi^{2s}}{(2 s)!}\sum_{j = 0}^{s}\binom{2s}{2j}\, \textrm{B} \bigg( \frac{2s+1}{2}, \frac{2s-2j+1}{2} \bigg)\, \textrm{B} \bigg(\frac{2s+1}{2}, \frac{2j+1}{2} \bigg)\\
& = \sum_{s=0}^{\infty} \frac{(-1)^{s}\, 2^{2s-1}\,\xi^{2s}}{(2 s)!} \frac{\pi ^{3/2}\, 16^{-s}\, (4 s)!\, \Gamma \left(s+\frac{1}{2}\right)}{(s!)^2\, (3 s)!} = \frac{1}{2} \pi ^2 \, _3F_4\left(\frac{1}{4},\frac{1}{2},\frac{3}{4};\frac{1}{3},\frac{2}{3},1,1;-\frac{16\, \xi ^2}{27} \right),
\end{alignat}
with the associated Fourier transform
\begin{alignat}{1}
\nonumber P_1\, (\Psi) &= \frac{2}{\pi^2} \int_{-\infty}^{\infty}
\left[ \frac{1}{2} \pi ^2 \, _3F_4\left(\frac{1}{4},\frac{1}{2},\frac{3}{4};\frac{1}{3},\frac{2}{3},1,1;-\frac{16\, \xi ^2}{27} \right)\right] \exp\, ( -\mathrm{i}\, \xi  \,\Psi \,)\, \frac{\mathrm{d}\xi}{2 \pi}\\
\label{eq:P1} &= 
 \begin{cases} 
 \vspace*{0.1cm}
 {\displaystyle
 \frac{3\,  G_{4,4}^{4,0}\left({\displaystyle\frac{27\, \Psi ^2}{64}} \bigg\vert
\begin{array}{c}
 -\frac{1}{6},\frac{1}{6},\frac{1}{2},\frac{1}{2} \\[0.25em]
 -\frac{1}{4},0,0,\frac{1}{4} \\
\end{array}
\right)}{4 \sqrt{2}\, \pi}};\quad &\mbox{for} \, \lvert \Psi \rvert < {\displaystyle \sqrt{\frac{64}{27}}}\\
0; &\mbox{otherwise},
\end{cases},
\end{alignat}
where $G_{p q}^{a b}$ stands for the Meijer G-function \cite{gradshteyn2014table}. Analogously,
\begin{equation}
\label{eq:P2}
\mathcal{I}_2 = 
\frac{8}{3}\, \mathrm{i}\, \xi  \, _4F_5\left(\frac{3}{4},1,1,\frac{5}{4};\frac{5}{6},\frac{7}{6},\frac{3}{2},\frac{3}{2},\frac{3}{2};-\frac{16\, \xi ^2}{27} \right),
\end{equation}
wherefore
\begin{alignat}{1}
\label{eq:IscP}
P\, (\Psi) &= \begin{cases} 
 \vspace*{0.1cm}
 {\displaystyle
 \frac{3\,  G_{4,4}^{4,0}\left({\displaystyle\frac{27\, \Psi ^2}{64}} \bigg\vert
\begin{array}{c}
 -\frac{1}{6},\frac{1}{6},\frac{1}{2},\frac{1}{2} \\[0.25em]
 -\frac{1}{4},0,0,\frac{1}{4} \\
\end{array}
\right)}{4 \sqrt{2}\, \pi}}\, \left[1 + \sgn\, (\Psi) \right]; \quad &\mbox{for} \, \lvert \Psi \rvert < {\displaystyle \sqrt{\frac{64}{27}}},\\
0; &\mbox{otherwise}.
\end{cases}
\end{alignat}
Even without explicitly calculating $\mathcal{I}_2$, this result could have been argued directly from Eq.~\eqref{eq:P1} as follows. Since $\Psi\, (x, y) > 0 \,\,\forall \,\, (x, y)$ in the ground state, the Fourier transform of $\mathcal{I}_2$ must necessarily cancel out any nonzero value of Eq.~\eqref{eq:P1} for $\Psi < 0$. What therefore remains to be determined is the form of FT$\,[\mathcal{I}_2]$ for $\Psi > 0$. At the same time, observe that $\int_\Psi P_1 (\Psi) = 1$. Altogether, $P\,(\Psi)$ itself must also respect the same normalization so the additional contribution from FT$\,[\mathcal{I}_2]$ cannot have any arbitrary functional form. The only possible function satisfying both these constraints is simply Eq.~\eqref{eq:P1} times the signum, which brings us to Eq.~\eqref{eq:IscP}.

\subsubsection{First excited state}

\begin{figure}[htb]
\centering
\subfigure[]{\label{fig:IscCF}
\includegraphics[width=0.475\textwidth]{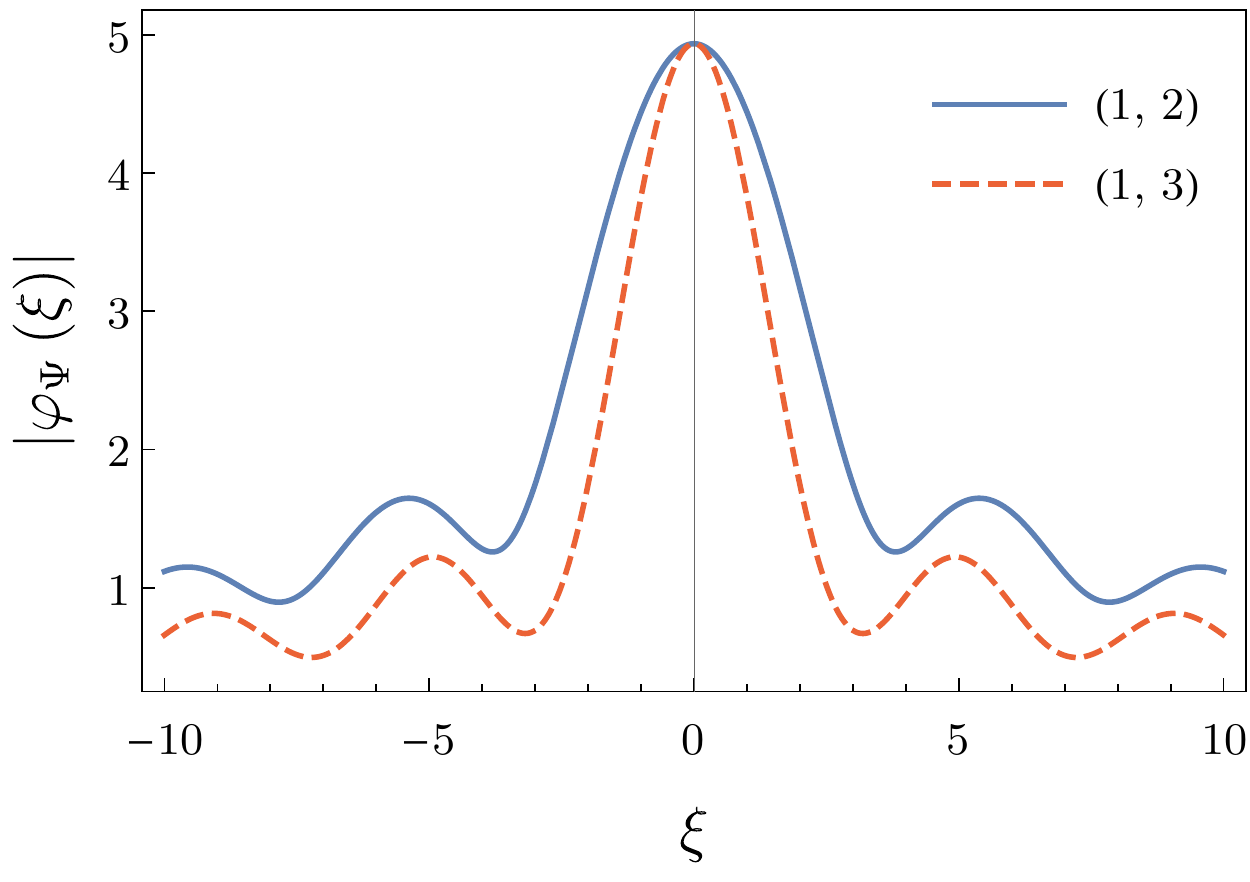}}
\subfigure[]{
\includegraphics[width=0.495\textwidth]{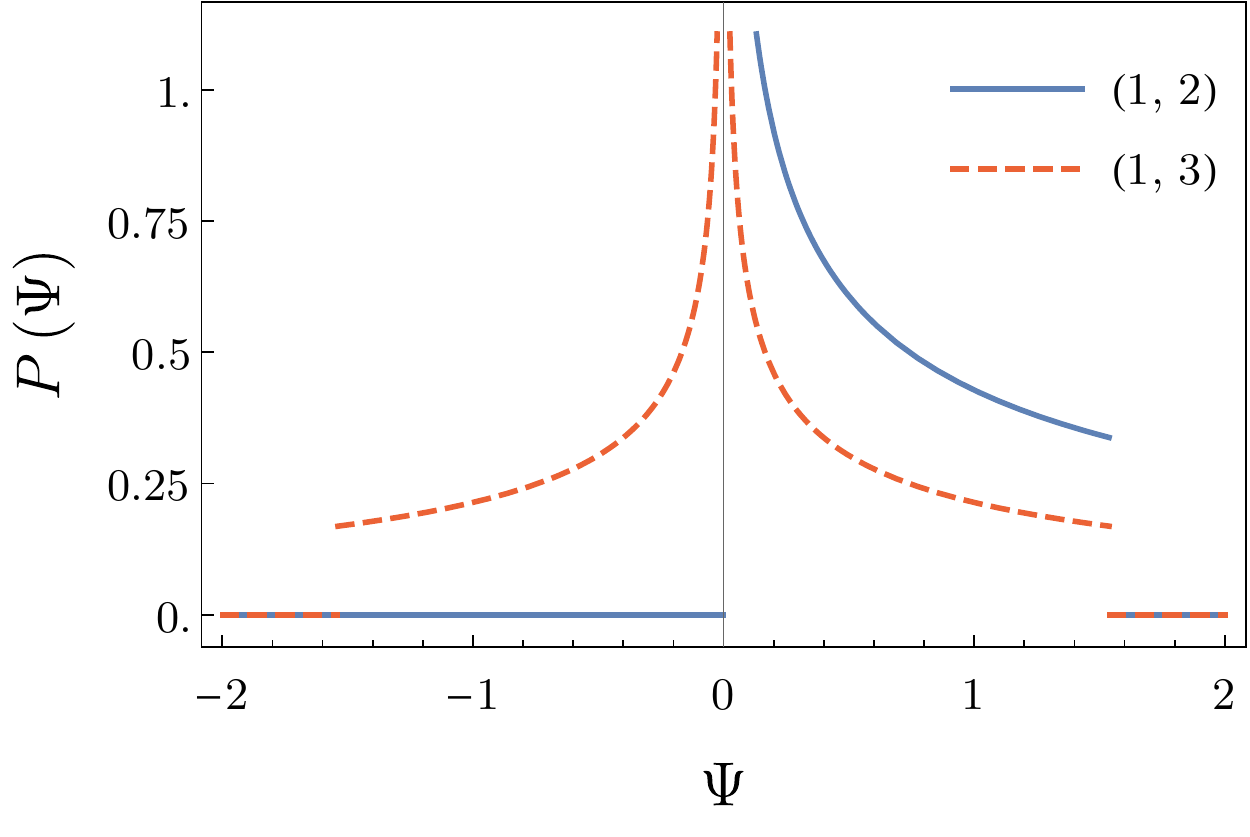}}
\caption{\label{fig:Isc}(a) The absolute value of the characteristic function, and (b) the probability distribution function of the wavefunction amplitudes for the two lowest states of the right-isosceles-triangular billiard, labelled by $(m, n)$. Note that the curves marked $(1,3)$ are identical to those for any eigenstate with quantum numbers $(d, 3d)$, for $d$ an even positive integer.}
\end{figure}

In the spirit of Eqs.~(\ref{eq:CFSplit1}, \ref{eq:CFSplit2}), the CF for the first excited state $(m, n) = (1,3)$ can be similarly decomposed into two polynomials of the wavefunction raised to even or odd powers. Fortunately, in this case, since $\Psi$ is odd under reflection about the line $y= \pi - x$ (the perpendicular bisector of the hypotenuse), all the odd terms vanish identically and $\mathcal{I}_2\,(\xi) = 0$. Proceeding as before, on integrating and resumming, we find
\begin{alignat}{1}
\label{eq:1Exc}
\varphi_{\Psi} \, (\xi) &= \frac{1}{2}\sum_{s=0}^{\infty} \frac{(-1)^{s}}{(2 s)!}\,\int_{0 }^{\pi }\int_{0 }^{\pi} \left( \sin\, x \, \sin\, 3 y - \sin\, 3 x \, \sin\, y \right)^{2s} \, \mathrm{d}x\,\mathrm{d}y\\
&= \frac{1}{2} \pi ^2 \, _3F_4\left(\frac{1}{4},\frac{1}{2},\frac{3}{4};\frac{1}{3},\frac{2}{3},1,1;-\frac{16\, \xi ^2}{27} \right),
\end{alignat}
and accordingly,
\begin{alignat}{1}
\label{eq:2Exc}
P\, (\Psi) &=  \begin{cases} 
 \vspace*{0.1cm}
 {\displaystyle
 \frac{3\,  G_{4,4}^{4,0}\left({\displaystyle\frac{27\, \Psi ^2}{64}} \bigg\vert
\begin{array}{c}
 -\frac{1}{6},\frac{1}{6},\frac{1}{2},\frac{1}{2} \\[0.25em]
 -\frac{1}{4},0,0,\frac{1}{4} \\
\end{array}
\right)}{4 \sqrt{2}\, \pi}};\quad &\mbox{for} \, \lvert \Psi \rvert < {\displaystyle \sqrt{\frac{64}{27}}},\\
0; &\mbox{otherwise}.
\end{cases}
\end{alignat}
The CFs (Eqs.~\ref{eq:P1}--\ref{eq:P2}, \ref{eq:1Exc}) and PDFs (Eqs.~\ref{eq:IscP}, \ref{eq:2Exc}) of the ground and excited states, respectively, are plotted and compared in Fig.~\ref{fig:Isc}.

\begin{figure}[htb]
\centering
\subfigure[]{
\includegraphics[width=0.25\textwidth]{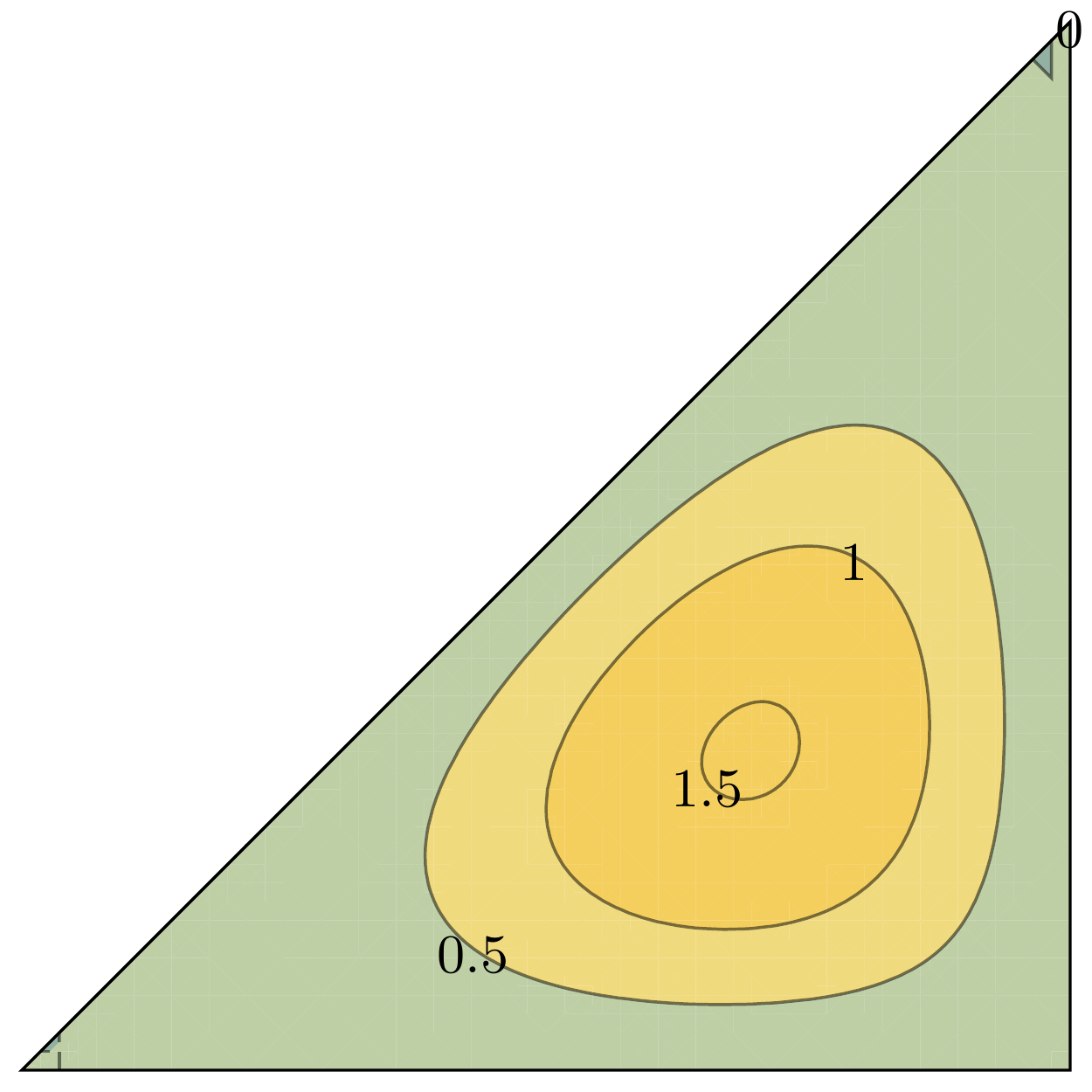}}\quad\quad
\subfigure[]{\label{Iscb}
\includegraphics[width=0.25\textwidth]{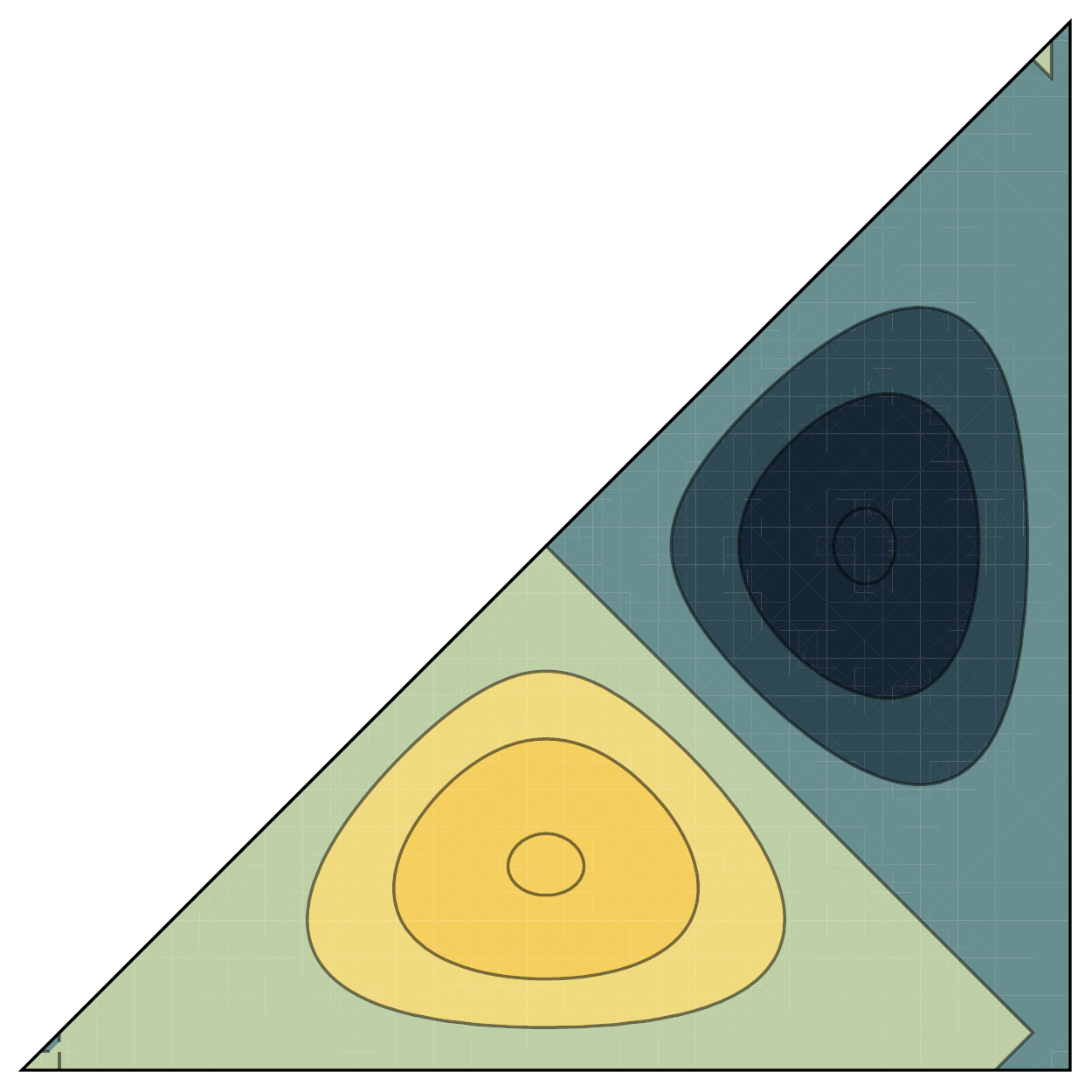}}\quad\quad
\subfigure[]{\label{Iscc}
\includegraphics[width=0.25\textwidth]{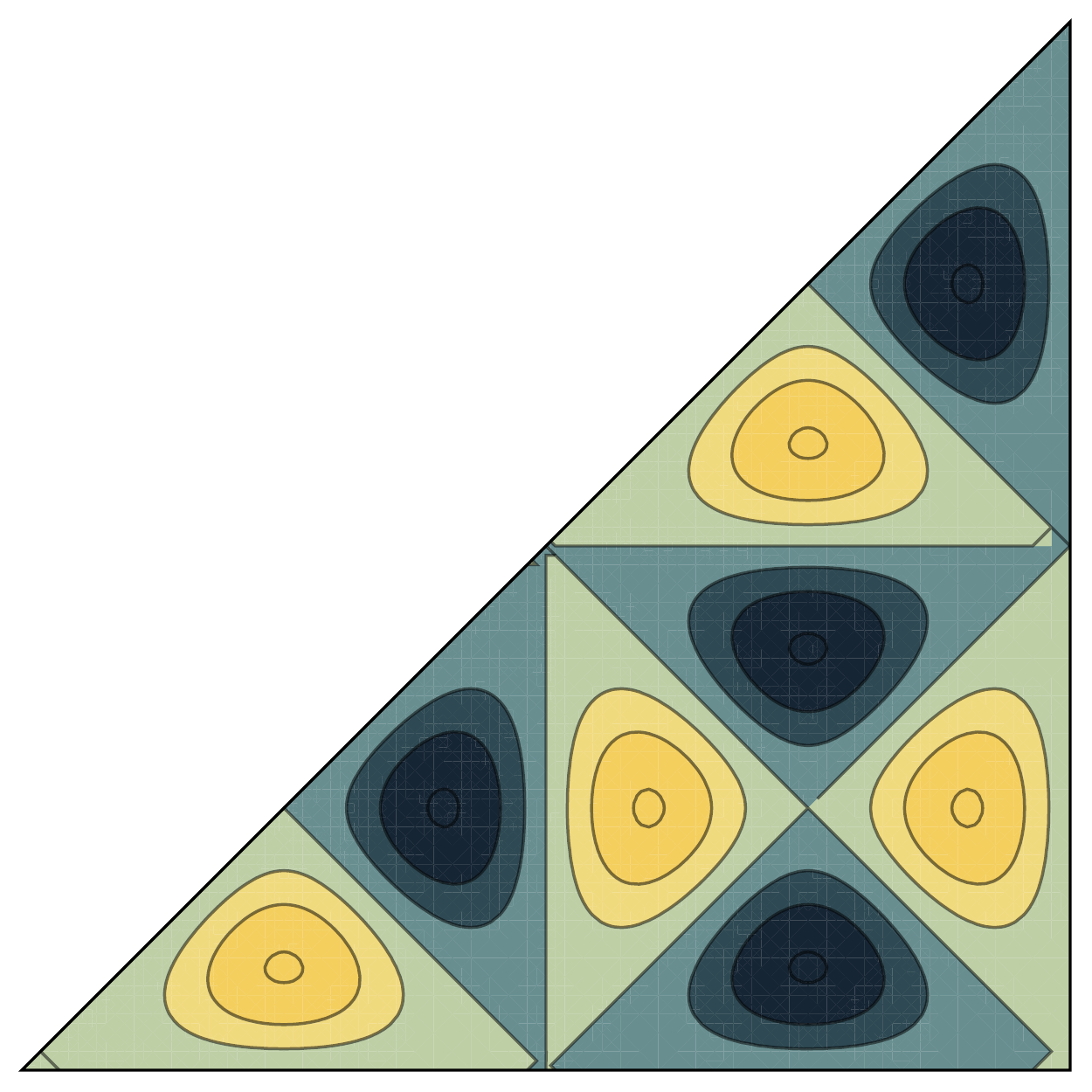}}\quad\quad
\subfigure{\includegraphics[height=1.50in, trim={0 .5cm 0 0}, clip]{ScaleRect.pdf}}
\caption{Contour plots of the wavefunctions of a quantum billiard in the shape of a right-angled isosceles triangle. (a) The eigenfunction of the ground state, $(1,2)$, is entirely positive. (b) The first excited state $(1,3)$ is antisymmetric about the altitude as $(1+3)\mbox{ mod }2 = 0$. (c) The wavefunction corresponding to $(2,6)$ is composed of $4 = 2^2$ repeated tiles---each of which is a replica of $(1,3)$.}
\end{figure}

Although algebraic intractability prevents the computation of the CF or the PDF for further excited states, a few general comments are in order. First, it is easy to observe that for \textsl{any} (positive) integer-valued $m$ and $n$
\begin{equation}
\label{eq:manual}
\int_{0 }^{\pi }\int_{0 }^{\pi}\left( \sin\, (m\,x)\, \sin\, (n\,y) - \sin\, (n\,x)\, \sin\, (m\,y) \right)^{t} \, \mathrm{d}x\,\mathrm{d}y = 
\begin{cases} 
\vspace*{0.1cm}\pi^2; \quad &\mbox{for } t = 0, \\
\displaystyle \frac{\pi^2}{2}; \quad &\mbox{for } t = 2.
\end{cases}
\end{equation}
Consequently, for any state in which all the odd terms in the expansion of the exponential vanish, the scale of oscillations of the CF, at least to second order in $\xi$, are universal and set by the two integrals above. Such states are precisely those that exhibit tiling \cite{aronovitch2012nodal} with an even number of tiles. The conditions for tiling are straightforward. For quantum numbers satisfying $(m + n) \mbox{ mod } 2 = 0$, the eigenfunction is antisymmetric about the line $y = \pi - x$ (Fig.~\ref{Iscb}) and forms two sub-triangles, or ``tiles''. Alternatively,  for $m,\,n$ such that $\gcd\,(m, n) = d > 1$, the wavefunction $\Psi_{m,n}$ consists of $d^2$ identical triangular copies of $\Psi_{m/d,n/d}$ (Fig.~\ref{Iscc}). Accordingly, Eq.~\eqref{eq:2Exc} correctly describes the PDF not only for $(1,3)$ but rather, for a whole hierarchy of states $\{(d, 3\,d) \, \vert \, d\mbox{ mod }2 = 0\}$, which extend to the semiclassical limit. Numerically calculating $\varphi_\Psi \,(\xi)$ for excited states confirms the expectation that the CF continues to structurally resemble Fig.~\ref{fig:IscCF} in its damped oscillatory character.

\subsection{Equilateral triangle}

\begin{figure}[htb]
\centering
\subfigure[]{\label{fig:EqCF}
\includegraphics[width=0.4825\textwidth]{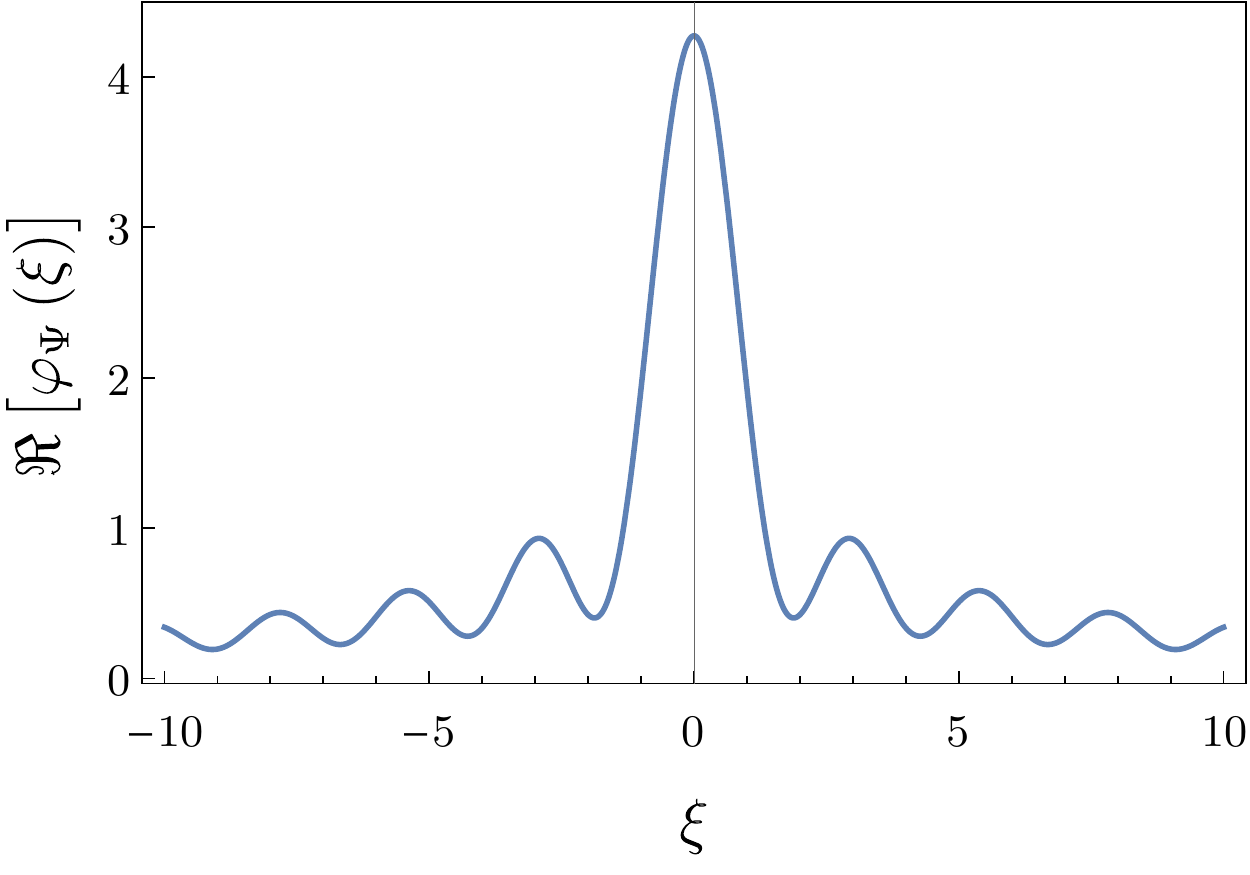}}
\subfigure[]{
\includegraphics[width=0.4925\textwidth]{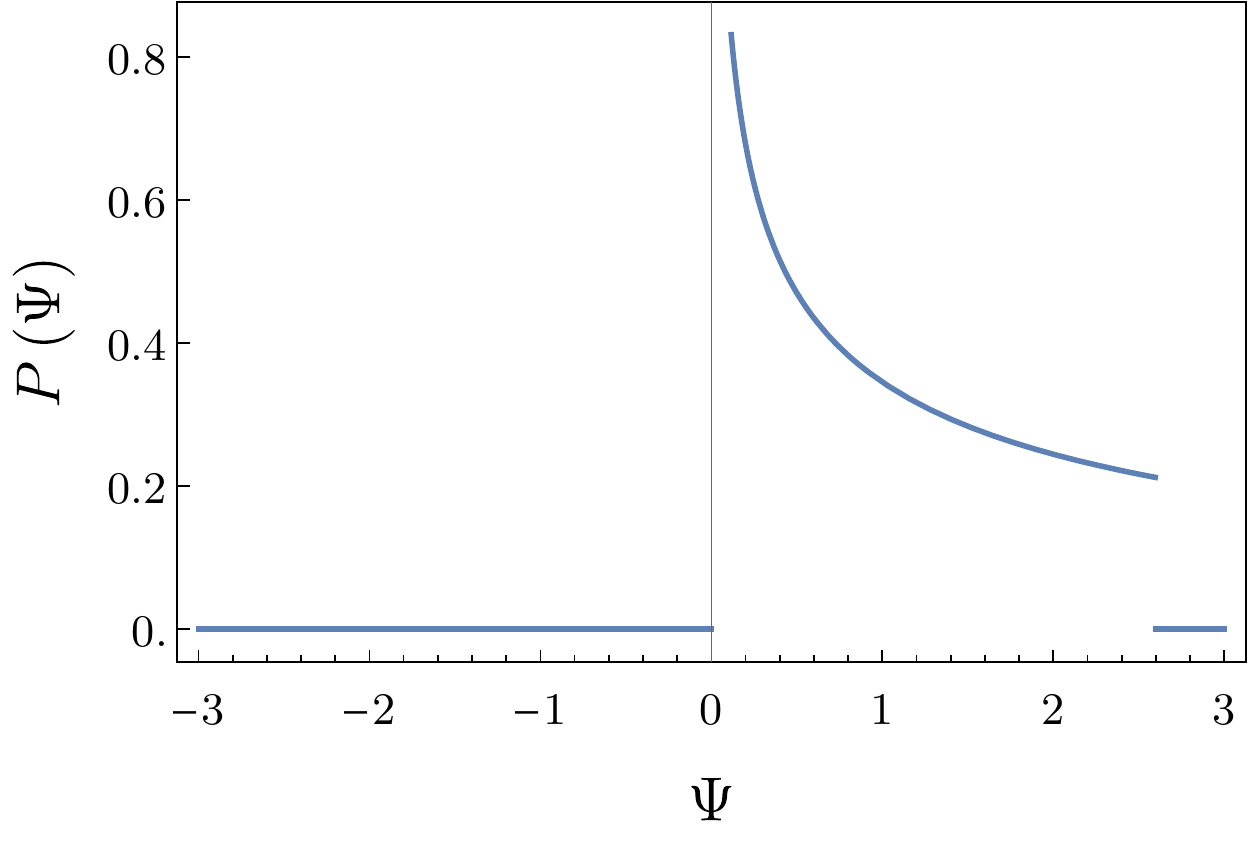}}
\caption{\label{fig:Eq}(a) The real part of the characteristic function (Eq.~\ref{eq:EqCF}) and (b) the probability distribution function of the wavefunction amplitudes (Eq.~\ref{eq:EqPDF}) for the ground state of the equilateral-triangular billiard.}
\end{figure}

Consider the equilateral-triangular domain of side $L = \pi$ and area $\mathcal{A} = \sqrt{3}\,\pi^2/4$: 
{
\begin{alignat}{1}
{\cal D} &= \bigg\{(x,y) \in \bigg[0,\frac{\pi}{2}\bigg] \times \bigg[0,\frac{\sqrt{3}\pi}{2}\bigg]: y\le\sqrt{3}x\bigg\} 
\cup \bigg\{(x,y) \in \bigg[\frac{\pi}{2},\pi \bigg] \times \bigg[0,\frac{\sqrt{3}\pi}{2}\bigg]: y\le\sqrt{3}(\pi-x)\bigg\}. 
\end{alignat}}
The Dirichlet eigenfunctions, which form a complete orthogonal basis, are \cite{brack2003semiclassical} 
\begin{alignat}{1}
\label{eq:wf}
&\Psi_{m,n}^{c,s} (x,y) =  \sin{\left((m-n)\frac{2\pi}{\sqrt{3}L}y\right)}(\cos, \sin)\left(-(m+n)\frac{2\pi}{3L}x\right)\\
&+\sin{\left(n\frac{2\pi}{\sqrt{3}L}y\right)} (\cos, \sin) \left((2m-n)\frac{2\pi}{3L}x \right) -\sin{\left(m\frac{2\pi}{\sqrt{3}L}y\right)} (\cos, \sin)\left((2n-m)\frac{2\pi}{3L}x\right)
\nonumber , 
\end{alignat}
where $m$ and $n$, as always, are integer quantum numbers with the restriction $m,n>0$. The eigenfunctions $\Psi_{m,n}^c$ and $\Psi_{m,n}^s$ correspond to the symmetric and antisymmetric modes respectively \cite{mccartin2003eigenstructure}. The eigenenergies of the Hamiltonian scale as $E_{m,n} = m^2+n^2-m\,n$. In the ground-state manifold ($m = 1,\, n = 2$), the CF is
\begin{equation}
\varphi_{\Psi} \, (\xi)\, = \int\int_{x, y \in \mathcal{D}} \exp \Bigg[ \,\mathrm{i}\, \xi  \, \left(\sin \left(\frac{4 y}{\sqrt{3}}\right)-2 \cos\, (2 x) \sin \left(\frac{2 y}{\sqrt{3}}\right) \right) \Bigg] \, \mathrm{d}y\,\mathrm{d}x. 
\end{equation}
As previously with the isosceles triangle, to avert having to integrate over the triangular domain, we extend the region of integration from $\mathcal{D}$ to the rhombus with vertices at $(0,0),\,(\pi,0),\, (\pi/2, \sqrt{3}\,\pi/2),$ and $(3\pi/2, \sqrt{3}\pi/2$). The terms bearing even powers of $\xi$ in the Taylor expansion of the exponential are unaffected by this tessellation. Next, we switch variables to the rhombic coordinate system defined by the transformation $u = x - y/\sqrt{3},\,v = 2 y/\sqrt{3}$. This has the twin advantage that, in terms of the transformed variables, the wavefunction reduces to the simple form
\begin{equation}
\label{eq:11}
\Psi \,(u,v) = 4 \sin\, u\,\sin\, v\, \sin\, (u+v),
\end{equation}
and the thusly simplified integration domain is just the square $[0, \pi]\times [0,\pi]$. Taking recourse to our previously-established repertoire of techniques, the integrals for the ground state can be computed exactly (we omit the details of the calculations here) to find 
\begin{equation}
\label{eq:EqCF}
\frac{1}{2}\sum_{s=0}^{\infty} \frac{(-1)^{s}}{(2 s)!}\,\int_{0 }^{\pi }\int_{0 }^{\pi}\frac{\sqrt{3}}{2} \left( 4 \sin\, u\,\sin\, v\, \sin\, (u+v)\right)^{2s} \, \mathrm{d}u\,\mathrm{d}v = \frac{\sqrt{3}}{4}  \pi ^2 \, _2F_3\left(\frac{1}{3},\frac{2}{3};\frac{1}{2},1,1;-\frac{27\, \xi ^2}{16} \right),
\end{equation}
and the concomitant PDF is
\begin{alignat}{1}
P(\Psi) = \begin{cases}
\label{eq:EqPDF}
\displaystyle
\vspace*{0.1cm}
\frac{1 + \sgn\, (\Psi)}{48\, \pi ^{5/2} \left \lvert \Psi \right \rvert ^{5/3}}\hspace*{-0.25cm} &\Bigg[\Gamma \left(\frac{2}{3}\right) \Gamma \left(-\frac{1}{6}\right) \left(\left(27-4 \Psi ^2\right) \, _2F_1\left(\frac{2}{3},\frac{2}{3};\frac{1}{3};\frac{4 \Psi ^2}{27}\right)-27 \, _2F_1\left(-\frac{1}{3},\frac{2}{3};\frac{1}{3};\frac{4 \Psi ^2}{27}\right)\right)\\
\displaystyle
\vspace*{0.1cm}
&+ \,72\, \Gamma \left(\frac{1}{3}\right) \Gamma \left(\frac{7}{6}\right) \left \lvert \Psi \right \rvert ^{4/3} \, _2F_1\left(\frac{1}{3},\frac{1}{3};\frac{2}{3};\frac{4 \Psi ^2}{27}\right)\Bigg];  \quad\mbox{for} \, \lvert \Psi \rvert < {\displaystyle \sqrt{\frac{27}{4}}},\\
0; &\mbox{otherwise.}
\end{cases}
\end{alignat}
The divergence of this distribution as $\Psi \rightarrow 0$ is indicated by Fig.~\ref{fig:Eq}; it is tempting to conjecture this density enhancement for small $\Psi$ to be characteristic of two-dimensional integrable billiards. Further evidence in support of this hypothesis, although purely statistical, comes from inspecting histograms constructed by numerically sampling the wavefunctions of the first few excited states, all of which seemingly bear out the ubiquity of the aforementioned singularity.

\begin{figure}[htb]
\centering
\subfigure[]{\label{fig:EqSc}
\includegraphics[width=0.485\textwidth]{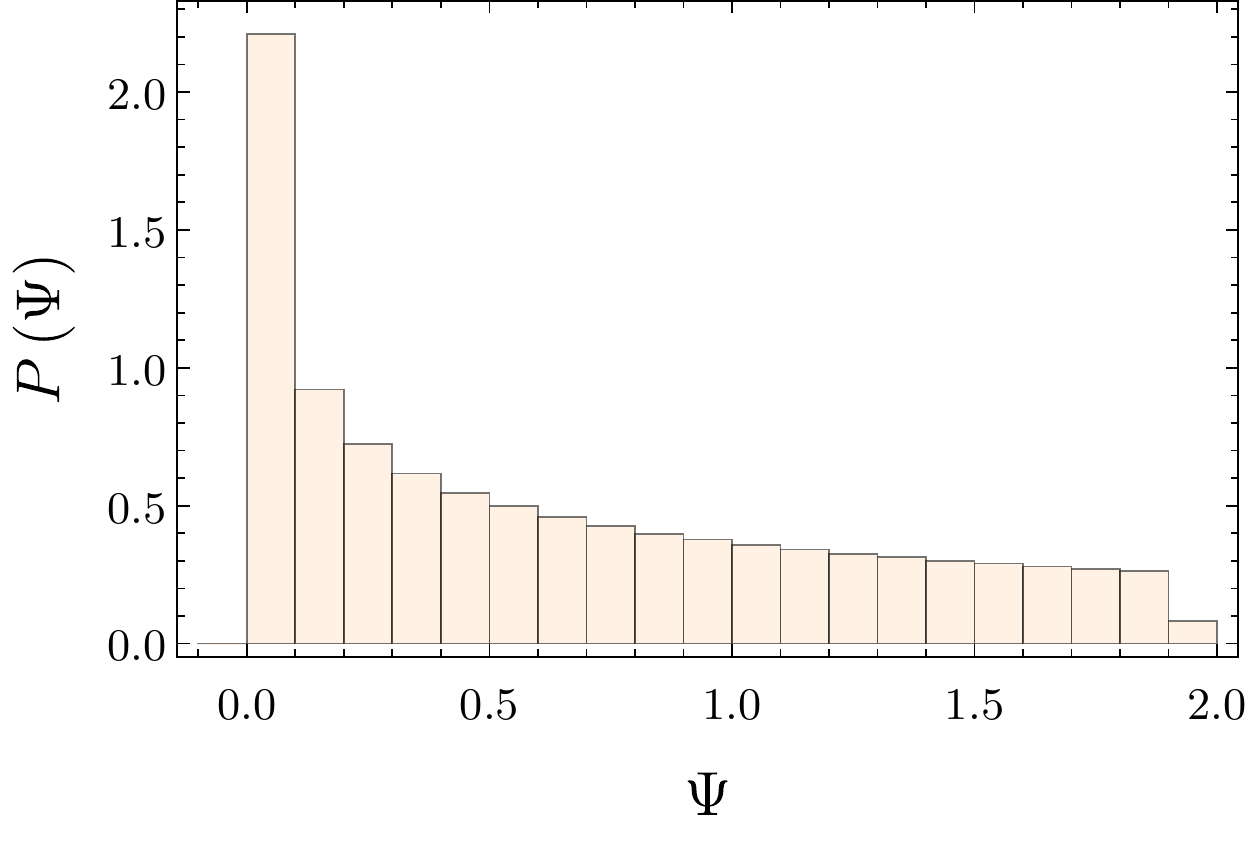}}
\subfigure[]{
\includegraphics[width=0.485\textwidth]{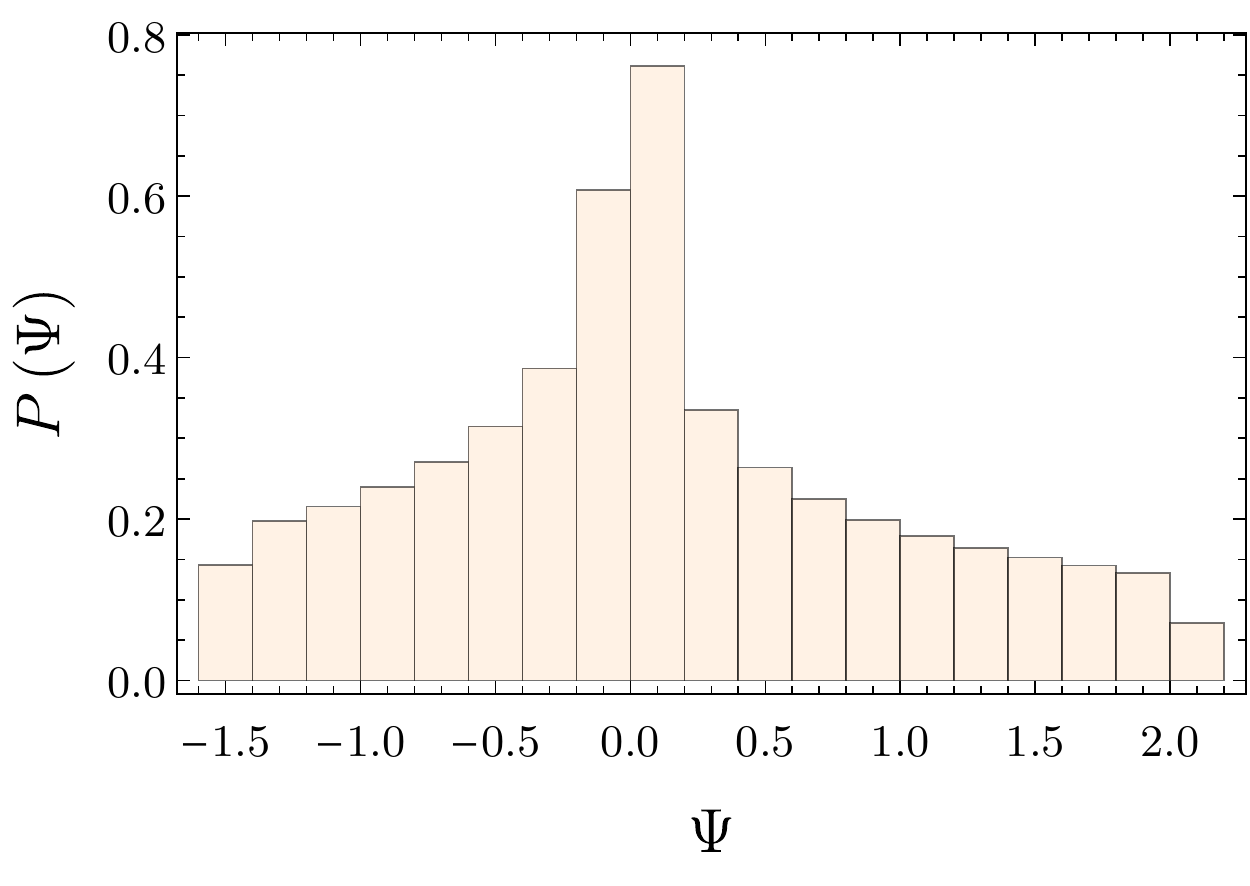}}
\caption{\label{fig:Sc} Histograms of wavefunction amplitudes for the (a) ground, and (b) first-excited state of the $30^{\circ}-60^{\circ}-90^{\circ}$ hemiequilateral triangle. The best-fit PDF in both cases is strictly convex and one ipso facto anticipates a singularity at $\Psi = 0$.}
\end{figure}

One could naively attempt to generalize the procedure above to analyze the excited states of the billiard. Unfortunately, what brings such well-intentioned efforts to grief is that the wavefunctions, for higher quantum numbers, are sums of three independent products of trigonometric functions, rather than just a solitary term as in Eq.~\eqref{eq:11}. This added complication calls for a trinomial expansion and resummation of the infinite series of coefficients in the analogue to Eq.~\eqref{eq:EqCF} no longer proves feasible. The same problem hounds even the ground-state wavefunction of the $30^{\circ}-60^{\circ}-90^{\circ}$ scalene-triangular billiard but, as corroborated by Fig.~\ref{fig:Sc}, our numerics once again point to a diverging PDF in the limit $\Psi \rightarrow 0$.

\section{Conclusion}

In this article, we have computed and compiled a novel set of exact distributions of wavefunction amplitudes for a host of integrable billiards---both separable and nonseparable. For the former category, the probability densities that we discover are especially useful because the dependence on the quantum numbers is weak. In general, the PDFs for integrable systems are found to be substantially different from (and perhaps, in some sense, mathematically richer than) the Gaussian amplitude distribution predicted by the random wave model. This distinction is seen to recurrently manifest itself in significantly enhanced probability densities at small amplitudes. The divergent terms can be formally characterized by expanding the distribution functions as power series in $\Psi$, whereby one finds, to leading order, 
\begin{alignat}{2} 
&\mbox{Square: }&& P\,(\Psi) \sim {\displaystyle \dfrac{\log (16/\Psi ^2)}{\pi ^2}}+\mathcal{O}\left(\Psi ^2\right), \\
& \mbox{Right-isosceles triangle: }&& P\,(\Psi) \sim \frac{\sqrt[4]{3}\, \pi ^{3/2}}{\Gamma \left(\frac{1}{12}\right) \Gamma \left(\frac{5}{12}\right) \Gamma \left(\frac{3}{4}\right)^4\,\sqrt{\Psi }} - \frac{\log \left(\Psi /16 \right)+1}{2 \pi ^2} +\mathcal{O} (\sqrt{\Psi} ),  \\
& \mbox{Equilateral triangle: }&& P\,(\Psi) \sim {\displaystyle -\frac{\Gamma \left(-\frac{2}{3}\right) \Gamma \left(\frac{7}{6}\right)}{\pi ^{5/2}\, \sqrt[3]{\Psi }}}+\mathcal{O}\left({\Psi ^{1/3}}\right).
\end{alignat}

It still remains a desideratum to express by equations the characteristic functions and distributions for the excited states of nonseparable billiards. Nonetheless, the importance of any exact results for such systems, even if they be for low-lying states, is underscored by multifarious considerations. First and foremost, as we have discussed, the distributions for the ground and first-excited states help shed light on several behavioral features that are presumably universal. Secondly, for ergodic billiards---in the semiclassical limit---it is believed that the chaotic nature of the underlying trajectories is unveiled only at an excitation via, for instance, a positive Lyapunov exponent. However, the low-lying states that \textsl{are} exactly known do not bear any elements of randomness whatsoever \cite{jain2002}, even as the Lyapunov exponents are positive for the corresponding energies. In light of this presumable insensitivity of the ground or first-excited states to the chaoticity/integrability of the billiard, analytical results thereon can be expected to have even more widespread applicability than general and hence, are of particularly nontrivial significance. Finally, the wavefunction of a given excited state, far from being independent of, is closely related to those for lower quantum numbers, as hinted by the intricate self-similarity in nodal patterns of triangular eigenfunctions \cite{samajdar2014JPA, samajdar2014nodal, manjunath2016difference, jain2017nodal}. Moreover, it has recently been demonstrated that within a class of eigenfunctions of an integrable, nonseparable billiard, all excited states can be built from the lowest one by repeated application of a raising operator \cite{mandwal2017}. This leads us to believe that the analysis and results presented here lay the foundations for future investigations into amplitude distribution functions in the time to come.

\section*{Acknowledgements}
 
RS is supported by the Purcell Fellowship.

\appendix

\section{Asymptotic distribution function for a circular billiard}
\label{app}
Another integrable billiard for which the amplitude distribution can be analytically calculated, albeit approximately, is the circle. A circular domain of radius $R$ in two dimensions, physically corresponding to a cylindrical infinite well, may be defined as $\mathcal{D} = \{(x, y): x^2 + y^2 \le R^2\}$. In polar coordinates, the solutions of the Schr\"{o}dinger equation (for a particle of mass $M$) are separable into angular and radial components and are given by \cite{Robinett1,0143-0807-24-3-302}
\begin{equation}
\Psi_{m,n} (r, \theta) = \frac{J_m (k_n\, r/R) \,\,\mathrm{e}^{\, \mathrm{i}\,m\, \theta}}{\sqrt{2 \pi \int_0^R \,\left[J_m (k_n\, r/R)\right]^2\, r\, \mathrm{d}r}}; \quad k = \sqrt{2 M E/\hbar},
\end{equation}
where $J_m (z)$ denotes the cylindrical Bessel function of the first kind, which has non-divergent solutions as $z \rightarrow 0$. The energy spectrum for the system is $E = \left[z_{m,n}\right]^2$, with $z_{m,n}$ representing the $n$th zero of the regular Bessel function $J_m (z)$, wherefore $k \sim z_{m,n}$. For convenience's sake, we consider a circle of radius $\pi$ (a choice, which, by virtue of scaling, is eventually irrelevant) and work with real solutions to the Schr\"{o}dinger equation. As outlined previously, we can work out the characteristic function 
\begin{alignat}{1}
\varphi_{\Psi} \, (\xi)\, &= \int_{0 }^{\pi }\hspace*{-0.1cm}\int_{0 }^{2 \pi } \exp \left[ \, \mathrm{i}\, \xi\, J_m (k_n\, r/\pi) \,\cos \,(m\,\theta)\, \right] \,r\,\mathrm{d}r\,\mathrm{d}\theta,
\end{alignat}
which, subject to the value of $m$, evaluates to
\begin{alignat}{1}
\varphi_{\Psi} \, (\xi)\, & = 
\begin{cases}
\vspace*{0.1cm} {\displaystyle \int_{0 }^{\pi } 2\pi \exp \left[\mathrm{i}\, \xi\, J_0 (k_n\, r/\pi)   \right] \,r\,\mathrm{d}r }; & m=0, \\
{\displaystyle \int_{0 }^{\pi } 2\pi J_0 \left( \left \lvert \xi\, J_m (k_n\, r/\pi) \right \rvert \right) \,r\,\mathrm{d}r }; & m \ne 0.
\end{cases}
\label{eq:13}
\end{alignat}
At this stage, however, Eq.~\eqref{eq:13} entails the integral over either the exponential of a Bessel function or nested Bessel functions, neither of which can be performed exactly. However, assuming sufficiently large quantum numbers ($m, n \gg 1$), it is reasonable to approximate the interior Bessel function by its large-order expansion \cite{olver1952some}:
\begin{equation}
J_\nu (z) \sim \frac{1}{\sqrt{2 \pi\,\nu}}\, \left(\frac{\mathrm{e}\, z}{2 \,\nu} \right)^\nu.
\end{equation}
Substituting this asymptotic form into Eq.~\eqref{eq:13}, after having converted $J_0$ into its power-series representation, we obtain
\begin{alignat}{1}
\nonumber\varphi_{\Psi} \, (\xi)\, &= \sum_{t=0}^{\infty} \, 2\pi\, \frac{(-1)^t}{t!\,\Gamma(t+1)} \, \bigg(\frac{\xi}{2}\bigg)^{2t}\, \frac{\mathrm{e}^{2 m t}\,k_n^{2 m t}}{(2\pi\,m)^{(2m+1) \,t}} \int_{0 }^{\pi } r^{2 m t + 1} \,\mathrm{d}r\\
&= \pi ^3 \, _1F_2\left(\frac{1}{m};1,1+\frac{1}{m};-\frac{2^{-2 m-3}\, \mathrm{e}^{2 m} \left( k_n / m\right)^{2 m} \,\xi ^2}{m\, \pi }\right),
\end{alignat}
whereupon a Fourier transform leads to
\begin{equation}
P \, (\Psi) =
\begin{cases}
{\displaystyle \frac{2^{\frac{1}{m}+2} m^{\frac{1}{m}+1} \pi ^{1/m}}{\mathrm{e}^2} \Bigg(\frac{\sqrt{\pi } \sec \left(\frac{\pi }{m}\right) \left \lvert \Psi \right \rvert ^{\frac{2}{m}-1} \left(k_n^{2 m}\right){}^{-1/m}}{\Gamma \left(\frac{1}{2}+\frac{1}{m}\right) \Gamma \left(\frac{m-1}{m}\right)}}&\\
{\displaystyle -\frac{\left(\Psi ^2 k_n^{-2 m}\right){}^{\frac{1}{m}-\frac{1}{2}} \mathrm{B}_{2^{2 m+1} \mathrm{e}^{-2 m} m^{2 m+1} \pi  \Psi ^2 k_n^{-2 m}}\left(\frac{1}{2}-\frac{1}{m},\frac{1}{2}\right)}{\pi  \sqrt{k_n^{2 m}}}\Bigg)}; &(\mathrm{e}\, k_n)^{2 m} >\pi \, (2 m)^{2 m+1}\, \Psi ^2,\\
0; &\mbox{ otherwise. }
\end{cases}
\end{equation}
Here, $\mathrm{B}_z (a, b)$ denotes the incomplete Euler beta function. While this distribution is expected to hold only in the manifold of highly excited states, the noteworthy feature is its limiting behavior for small amplitudes. As $\Psi \rightarrow 0$, it diverges as
\begin{equation}
P\,(\Psi) \sim \left( \frac{2^{\frac{1}{m}+3}\, m^{\frac{1}{m}+2}\, \pi ^{\frac{1}{m}+\frac{5}{2}} \sec \left(\frac{\pi }{m}\right) \left(k_n^{2 m}\right){}^{-1/m}}{\mathrm{e}^2\, \Gamma \left(\frac{1}{2}+\frac{1}{m}\right) \Gamma \left(-\frac{1}{m}\right)} \right) \, \frac{1}{\lvert \Psi \rvert ^{1 -\frac{2}{m}} } + \mathcal{C}_{m,n} + \mathcal{O}\, (\Psi^2)
\end{equation}
(since $m > 2$), with $\mathcal{C}_{m,n}$ some quantum number-dependent constant. This singularity is of a power-law nature, as opposed to the logarithmic dependence noted for the rectangle. The complementary regime of $\Psi \gtrsim 1$ is analyzed by Ref.~\onlinecite{beugeling2017statistical}, which specifically addresses the tails of the distribution.\\

\bibliography{DensityBib.bib}

\begin{thebibliography}{46}%
\makeatletter
\providecommand \@ifxundefined [1]{%
 \@ifx{#1\undefined}
}%
\providecommand \@ifnum [1]{%
 \ifnum #1\expandafter \@firstoftwo
 \else \expandafter \@secondoftwo
 \fi
}%
\providecommand \@ifx [1]{%
 \ifx #1\expandafter \@firstoftwo
 \else \expandafter \@secondoftwo
 \fi
}%
\providecommand \natexlab [1]{#1}%
\providecommand \enquote  [1]{``#1''}%
\providecommand \bibnamefont  [1]{#1}%
\providecommand \bibfnamefont [1]{#1}%
\providecommand \citenamefont [1]{#1}%
\providecommand \href@noop [0]{\@secondoftwo}%
\providecommand \href [0]{\begingroup \@sanitize@url \@href}%
\providecommand \@href[1]{\@@startlink{#1}\@@href}%
\providecommand \@@href[1]{\endgroup#1\@@endlink}%
\providecommand \@sanitize@url [0]{\catcode `\\12\catcode `\$12\catcode
  `\&12\catcode `\#12\catcode `\^12\catcode `\_12\catcode `\%12\relax}%
\providecommand \@@startlink[1]{}%
\providecommand \@@endlink[0]{}%
\providecommand \url  [0]{\begingroup\@sanitize@url \@url }%
\providecommand \@url [1]{\endgroup\@href {#1}{\urlprefix }}%
\providecommand \urlprefix  [0]{URL }%
\providecommand \Eprint [0]{\href }%
\providecommand \doibase [0]{http://dx.doi.org/}%
\providecommand \selectlanguage [0]{\@gobble}%
\providecommand \bibinfo  [0]{\@secondoftwo}%
\providecommand \bibfield  [0]{\@secondoftwo}%
\providecommand \translation [1]{[#1]}%
\providecommand \BibitemOpen [0]{}%
\providecommand \bibitemStop [0]{}%
\providecommand \bibitemNoStop [0]{.\EOS\space}%
\providecommand \EOS [0]{\spacefactor3000\relax}%
\providecommand \BibitemShut  [1]{\csname bibitem#1\endcsname}%
\let\auto@bib@innerbib\@empty
\bibitem [{\citenamefont {Percival}(1973)}]{percival1973regular}%
  \BibitemOpen
  \bibfield  {author} {\bibinfo {author} {\bibfnamefont {I.~C.}\ \bibnamefont
  {Percival}},\ }\bibfield  {title} {\enquote {\bibinfo {title} {Regular and
  irregular spectra},}\ }\href {\doibase 10.1088/0022-3700/6/9/002} {\bibfield
  {journal} {\bibinfo  {journal} {J. Phys. B: At. Mol. Phys.}\ }\textbf
  {\bibinfo {volume} {6}},\ \bibinfo {pages} {L229} (\bibinfo {year}
  {1973})}\BibitemShut {NoStop}%
\bibitem [{\citenamefont {Berry}(1977)}]{berry1977regular}%
  \BibitemOpen
  \bibfield  {author} {\bibinfo {author} {\bibfnamefont {M.~V.}\ \bibnamefont
  {Berry}},\ }\bibfield  {title} {\enquote {\bibinfo {title} {Regular and
  irregular semiclassical wavefunctions},}\ }\href {\doibase
  10.1088/0305-4470/10/12/016} {\bibfield  {journal} {\bibinfo  {journal} {J.
  Phys. A: Math. Gen.}\ }\textbf {\bibinfo {volume} {10}},\ \bibinfo {pages}
  {2083} (\bibinfo {year} {1977})}\BibitemShut {NoStop}%
\bibitem [{\citenamefont {Berry}(1983)}]{berry1983semiclassical}%
  \BibitemOpen
  \bibfield  {author} {\bibinfo {author} {\bibfnamefont {M.~V.}\ \bibnamefont
  {Berry}},\ }\bibfield  {title} {\enquote {\bibinfo {title} {Semiclassical
  mechanics of regular and irregular motion},}\ }in\ \href@noop {} {\emph
  {\bibinfo {booktitle} {Chaotic Behaviour of Deterministic Systems}}},\
  Vol.~\bibinfo {volume} {36},\ \bibinfo {editor} {edited by\ \bibinfo {editor}
  {\bibfnamefont {G.}~\bibnamefont {Iooss}}, \bibinfo {editor} {\bibfnamefont
  {R.~H.~G.}\ \bibnamefont {Hellemann}}, \ and\ \bibinfo {editor}
  {\bibfnamefont {R.}~\bibnamefont {Stora}}},\ \bibinfo {organization} {Les
  Houches lectures}\ (\bibinfo  {publisher} {North-Holland Amsterdam},\
  \bibinfo {year} {1983})\ pp.\ \bibinfo {pages} {171--271}\BibitemShut
  {NoStop}%
\bibitem [{\citenamefont {B\"{a}cker}(2007)}]{phdthesis}%
  \BibitemOpen
  \bibfield  {author} {\bibinfo {author} {\bibfnamefont {A.}~\bibnamefont
  {B\"{a}cker}},\ }\emph {\bibinfo {title} {Eigenfunctions in chaotic quantum
  systems}},\ \href
  {http://www.qucosa.de/fileadmin/data/qucosa/documents/159/1213275874643-5042.pdf}
  {Ph.D. thesis},\ \bibinfo  {school} {Technische Universit\"{a}t Dresden}
  (\bibinfo {year} {2007})\BibitemShut {NoStop}%
\bibitem [{\citenamefont {Shapiro}\ and\ \citenamefont
  {Goelman}(1984)}]{shapiro1984onset}%
  \BibitemOpen
  \bibfield  {author} {\bibinfo {author} {\bibfnamefont {M.}~\bibnamefont
  {Shapiro}}\ and\ \bibinfo {author} {\bibfnamefont {G.}~\bibnamefont
  {Goelman}},\ }\bibfield  {title} {\enquote {\bibinfo {title} {Onset of chaos
  in an isolated energy eigenstate},}\ }\href {\doibase
  10.1103/PhysRevLett.53.1714} {\bibfield  {journal} {\bibinfo  {journal}
  {Phys. Rev. Lett.}\ }\textbf {\bibinfo {volume} {53}},\ \bibinfo {pages}
  {1714} (\bibinfo {year} {1984})}\BibitemShut {NoStop}%
\bibitem [{\citenamefont {McDonald}\ and\ \citenamefont
  {Kaufman}(1988)}]{mcdonald1988wave}%
  \BibitemOpen
  \bibfield  {author} {\bibinfo {author} {\bibfnamefont {S.~W.}\ \bibnamefont
  {McDonald}}\ and\ \bibinfo {author} {\bibfnamefont {A.~N.}\ \bibnamefont
  {Kaufman}},\ }\bibfield  {title} {\enquote {\bibinfo {title} {Wave chaos in
  the stadium: statistical properties of short-wave solutions of the
  {H}elmholtz equation},}\ }\href {\doibase 10.1103/PhysRevA.37.3067}
  {\bibfield  {journal} {\bibinfo  {journal} {Phys. Rev. A}\ }\textbf {\bibinfo
  {volume} {37}},\ \bibinfo {pages} {3067} (\bibinfo {year}
  {1988})}\BibitemShut {NoStop}%
\bibitem [{\citenamefont {O'connor}\ and\ \citenamefont
  {Heller}(1988)}]{o1988quantum}%
  \BibitemOpen
  \bibfield  {author} {\bibinfo {author} {\bibfnamefont {P.~W.}\ \bibnamefont
  {O'connor}}\ and\ \bibinfo {author} {\bibfnamefont {E.~J.}\ \bibnamefont
  {Heller}},\ }\bibfield  {title} {\enquote {\bibinfo {title} {Quantum
  localization for a strongly classically chaotic system},}\ }\href {\doibase
  10.1103/PhysRevLett.61.2288} {\bibfield  {journal} {\bibinfo  {journal}
  {Phys. Rev. Lett.}\ }\textbf {\bibinfo {volume} {61}},\ \bibinfo {pages}
  {2288} (\bibinfo {year} {1988})}\BibitemShut {NoStop}%
\bibitem [{\citenamefont {Aurich}\ and\ \citenamefont
  {Steiner}(1991)}]{aurich1991exact}%
  \BibitemOpen
  \bibfield  {author} {\bibinfo {author} {\bibfnamefont {R.}~\bibnamefont
  {Aurich}}\ and\ \bibinfo {author} {\bibfnamefont {F.}~\bibnamefont
  {Steiner}},\ }\bibfield  {title} {\enquote {\bibinfo {title} {Exact theory
  for the quantum eigenstates of a strongly chaotic system},}\ }\href {\doibase
  10.1016/0167-2789(91)90097-S} {\bibfield  {journal} {\bibinfo  {journal}
  {Physica D}\ }\textbf {\bibinfo {volume} {48}},\ \bibinfo {pages} {445--470}
  (\bibinfo {year} {1991})}\BibitemShut {NoStop}%
\bibitem [{\citenamefont {Hejhal}\ and\ \citenamefont
  {Rackner}(1992)}]{hejhal1992topography}%
  \BibitemOpen
  \bibfield  {author} {\bibinfo {author} {\bibfnamefont {D.~A.}\ \bibnamefont
  {Hejhal}}\ and\ \bibinfo {author} {\bibfnamefont {B.~N.}\ \bibnamefont
  {Rackner}},\ }\bibfield  {title} {\enquote {\bibinfo {title} {On the
  topography of {M}aass waveforms for {PSL} (2, $\mathbb{Z}$)},}\ }\href
  {\doibase 10.1080/10586458.1992.10504562} {\bibfield  {journal} {\bibinfo
  {journal} {Exp. Math.}\ }\textbf {\bibinfo {volume} {1}},\ \bibinfo {pages}
  {275--305} (\bibinfo {year} {1992})}\BibitemShut {NoStop}%
\bibitem [{\citenamefont {Aurich}\ and\ \citenamefont
  {Steiner}(1993)}]{aurich1993statistical}%
  \BibitemOpen
  \bibfield  {author} {\bibinfo {author} {\bibfnamefont {R.}~\bibnamefont
  {Aurich}}\ and\ \bibinfo {author} {\bibfnamefont {F.}~\bibnamefont
  {Steiner}},\ }\bibfield  {title} {\enquote {\bibinfo {title} {Statistical
  properties of highly excited quantum eigenstates of a strongly chaotic
  system},}\ }\href {\doibase 10.1016/0167-2789(93)90255-Y} {\bibfield
  {journal} {\bibinfo  {journal} {Physica D}\ }\textbf {\bibinfo {volume}
  {64}},\ \bibinfo {pages} {185--214} (\bibinfo {year} {1993})}\BibitemShut
  {NoStop}%
\bibitem [{\citenamefont {Li}\ and\ \citenamefont
  {Robnik}(1994)}]{li1994statistical}%
  \BibitemOpen
  \bibfield  {author} {\bibinfo {author} {\bibfnamefont {B.}~\bibnamefont
  {Li}}\ and\ \bibinfo {author} {\bibfnamefont {M.}~\bibnamefont {Robnik}},\
  }\bibfield  {title} {\enquote {\bibinfo {title} {Statistical properties of
  high-lying chaotic eigenstates},}\ }\href {\doibase
  10.1088/0305-4470/27/16/017} {\bibfield  {journal} {\bibinfo  {journal} {J.
  Phys. A: Math. Gen.}\ }\textbf {\bibinfo {volume} {27}},\ \bibinfo {pages}
  {5509} (\bibinfo {year} {1994})}\BibitemShut {NoStop}%
\bibitem [{\citenamefont {Jalabert}, \citenamefont {Stone},\ and\ \citenamefont
  {Alhassid}(1992)}]{PhysRevLett.68.3468}%
  \BibitemOpen
  \bibfield  {author} {\bibinfo {author} {\bibfnamefont {R.~A.}\ \bibnamefont
  {Jalabert}}, \bibinfo {author} {\bibfnamefont {A.~D.}\ \bibnamefont {Stone}},
  \ and\ \bibinfo {author} {\bibfnamefont {Y.}~\bibnamefont {Alhassid}},\
  }\bibfield  {title} {\enquote {\bibinfo {title} {Statistical theory of
  {C}oulomb blockade oscillations: Quantum chaos in quantum dots},}\ }\href
  {\doibase 10.1103/PhysRevLett.68.3468} {\bibfield  {journal} {\bibinfo
  {journal} {Phys. Rev. Lett.}\ }\textbf {\bibinfo {volume} {68}},\ \bibinfo
  {pages} {3468--3471} (\bibinfo {year} {1992})}\BibitemShut {NoStop}%
\bibitem [{\citenamefont {Gr\'emaud}, \citenamefont {Delande},\ and\
  \citenamefont {Gay}(1993)}]{PhysRevLett.70.1615}%
  \BibitemOpen
  \bibfield  {author} {\bibinfo {author} {\bibfnamefont {B.}~\bibnamefont
  {Gr\'emaud}}, \bibinfo {author} {\bibfnamefont {D.}~\bibnamefont {Delande}},
  \ and\ \bibinfo {author} {\bibfnamefont {J.~C.}\ \bibnamefont {Gay}},\
  }\bibfield  {title} {\enquote {\bibinfo {title} {Origin of narrow resonances
  in the diamagnetic {R}ydberg spectrum},}\ }\href {\doibase
  10.1103/PhysRevLett.70.1615} {\bibfield  {journal} {\bibinfo  {journal}
  {Phys. Rev. Lett.}\ }\textbf {\bibinfo {volume} {70}},\ \bibinfo {pages}
  {1615--1618} (\bibinfo {year} {1993})}\BibitemShut {NoStop}%
\bibitem [{\citenamefont {Mirlin}\ and\ \citenamefont
  {Fyodorov}(1993)}]{mirlin1993statistics}%
  \BibitemOpen
  \bibfield  {author} {\bibinfo {author} {\bibfnamefont {A.~D.}\ \bibnamefont
  {Mirlin}}\ and\ \bibinfo {author} {\bibfnamefont {Y.~V.}\ \bibnamefont
  {Fyodorov}},\ }\bibfield  {title} {\enquote {\bibinfo {title} {The statistics
  of eigenvector components of random band matrices: analytical results},}\
  }\href {\doibase 10.1088/0305-4470/26/12/012} {\bibfield  {journal} {\bibinfo
   {journal} {J. Phys. A: Math. Gen.}\ }\textbf {\bibinfo {volume} {26}},\
  \bibinfo {pages} {L551} (\bibinfo {year} {1993})}\BibitemShut {NoStop}%
\bibitem [{\citenamefont {Efetov}\ and\ \citenamefont
  {Prigodin}(1993)}]{efetov1993local}%
  \BibitemOpen
  \bibfield  {author} {\bibinfo {author} {\bibfnamefont {K.~B.}\ \bibnamefont
  {Efetov}}\ and\ \bibinfo {author} {\bibfnamefont {V.~N.}\ \bibnamefont
  {Prigodin}},\ }\bibfield  {title} {\enquote {\bibinfo {title} {Local density
  of states distribution and {NMR} in small metallic particles},}\ }\href
  {\doibase 10.1103/PhysRevLett.70.1315} {\bibfield  {journal} {\bibinfo
  {journal} {Phys. Rev. Lett.}\ }\textbf {\bibinfo {volume} {70}},\ \bibinfo
  {pages} {1315} (\bibinfo {year} {1993})}\BibitemShut {NoStop}%
\bibitem [{\citenamefont {Fyodorov}\ and\ \citenamefont
  {Mirlin}(1995)}]{fyodorov1995mesoscopic}%
  \BibitemOpen
  \bibfield  {author} {\bibinfo {author} {\bibfnamefont {Y.~V.}\ \bibnamefont
  {Fyodorov}}\ and\ \bibinfo {author} {\bibfnamefont {A.~D.}\ \bibnamefont
  {Mirlin}},\ }\bibfield  {title} {\enquote {\bibinfo {title} {Mesoscopic
  fluctuations of eigenfunctions and level-velocity distribution in disordered
  metals},}\ }\href {\doibase 10.1103/PhysRevB.51.13403} {\bibfield  {journal}
  {\bibinfo  {journal} {Phys. Rev. B}\ }\textbf {\bibinfo {volume} {51}},\
  \bibinfo {pages} {13403} (\bibinfo {year} {1995})}\BibitemShut {NoStop}%
\bibitem [{\citenamefont {Hlushchuk}\ \emph {et~al.}(2001)\citenamefont
  {Hlushchuk}, \citenamefont {Sirko}, \citenamefont {Kuhl}, \citenamefont
  {Barth},\ and\ \citenamefont {St\"ockmann}}]{PhysRevE.63.046208}%
  \BibitemOpen
  \bibfield  {author} {\bibinfo {author} {\bibfnamefont {Y.}~\bibnamefont
  {Hlushchuk}}, \bibinfo {author} {\bibfnamefont {L.}~\bibnamefont {Sirko}},
  \bibinfo {author} {\bibfnamefont {U.}~\bibnamefont {Kuhl}}, \bibinfo {author}
  {\bibfnamefont {M.}~\bibnamefont {Barth}}, \ and\ \bibinfo {author}
  {\bibfnamefont {H.-J.}\ \bibnamefont {St\"ockmann}},\ }\bibfield  {title}
  {\enquote {\bibinfo {title} {Experimental investigation of a regime of
  {W}igner ergodicity in microwave rough billiards},}\ }\href {\doibase
  10.1103/PhysRevE.63.046208} {\bibfield  {journal} {\bibinfo  {journal} {Phys.
  Rev. E}\ }\textbf {\bibinfo {volume} {63}},\ \bibinfo {pages} {046208}
  (\bibinfo {year} {2001})}\BibitemShut {NoStop}%
\bibitem [{\citenamefont {Markus}\ and\ \citenamefont
  {Meyer}(1974)}]{markus1974generic}%
  \BibitemOpen
  \bibfield  {author} {\bibinfo {author} {\bibfnamefont {L.}~\bibnamefont
  {Markus}}\ and\ \bibinfo {author} {\bibfnamefont {K.~R.}\ \bibnamefont
  {Meyer}},\ }\bibfield  {title} {\enquote {\bibinfo {title} {Generic
  {H}amiltonian dynamical systems are neither integrable nor ergodic},}\ }in\
  \href {\doibase 10.1090/memo/0144} {\emph {\bibinfo {booktitle} {Mem. Amer.
  Math. Soc.}}},\ Vol.\ \bibinfo {volume} {144}\ (\bibinfo  {publisher}
  {American Mathematical Society},\ \bibinfo {address} {Providence, RI},\
  \bibinfo {year} {1974})\BibitemShut {NoStop}%
\bibitem [{\citenamefont {B\"{a}cker}\ and\ \citenamefont
  {Schubert}(2002)}]{0305-4470-35-3-306}%
  \BibitemOpen
  \bibfield  {author} {\bibinfo {author} {\bibfnamefont {A.}~\bibnamefont
  {B\"{a}cker}}\ and\ \bibinfo {author} {\bibfnamefont {R.}~\bibnamefont
  {Schubert}},\ }\bibfield  {title} {\enquote {\bibinfo {title} {Amplitude
  distribution of eigenfunctions in mixed systems},}\ }\href {\doibase
  10.1088/0305-4470/35/3/306} {\bibfield  {journal} {\bibinfo  {journal} {J.
  Phys. A: Math. Gen.}\ }\textbf {\bibinfo {volume} {35}},\ \bibinfo {pages}
  {527} (\bibinfo {year} {2002})}\BibitemShut {NoStop}%
\bibitem [{\citenamefont {Biswas}\ and\ \citenamefont {Jain}(1990)}]{piby3}%
  \BibitemOpen
  \bibfield  {author} {\bibinfo {author} {\bibfnamefont {D.}~\bibnamefont
  {Biswas}}\ and\ \bibinfo {author} {\bibfnamefont {S.~R.}\ \bibnamefont
  {Jain}},\ }\bibfield  {title} {\enquote {\bibinfo {title} {Quantum
  description of a pseudointegrable system: The $\pi$/3-rhombus billiard},}\
  }\href {\doibase 10.1103/PhysRevA.42.3170} {\bibfield  {journal} {\bibinfo
  {journal} {Phys. Rev. A}\ }\textbf {\bibinfo {volume} {42}},\ \bibinfo
  {pages} {3170} (\bibinfo {year} {1990})}\BibitemShut {NoStop}%
\bibitem [{\citenamefont {Efetov}(1983)}]{efetov1983supersymmetry}%
  \BibitemOpen
  \bibfield  {author} {\bibinfo {author} {\bibfnamefont {K.~B.}\ \bibnamefont
  {Efetov}},\ }\bibfield  {title} {\enquote {\bibinfo {title} {Supersymmetry
  and theory of disordered metals},}\ }\href {\doibase
  10.1080/00018738300101531} {\bibfield  {journal} {\bibinfo  {journal} {Adv.
  Phys.}\ }\textbf {\bibinfo {volume} {32}},\ \bibinfo {pages} {53--127}
  (\bibinfo {year} {1983})}\BibitemShut {NoStop}%
\bibitem [{\citenamefont {Verbaarschot}, \citenamefont {Weidenm{\"u}ller},\
  and\ \citenamefont {Zirnbauer}(1985)}]{verbaarschot1985grassmann}%
  \BibitemOpen
  \bibfield  {author} {\bibinfo {author} {\bibfnamefont {J.~J.~M.}\
  \bibnamefont {Verbaarschot}}, \bibinfo {author} {\bibfnamefont {H.~A.}\
  \bibnamefont {Weidenm{\"u}ller}}, \ and\ \bibinfo {author} {\bibfnamefont
  {M.~R.}\ \bibnamefont {Zirnbauer}},\ }\bibfield  {title} {\enquote {\bibinfo
  {title} {Grassmann integration in stochastic quantum physics: the case of
  compound-nucleus scattering},}\ }\href {\doibase
  10.1016/0370-1573(85)90070-5} {\bibfield  {journal} {\bibinfo  {journal}
  {Phys. Rep.}\ }\textbf {\bibinfo {volume} {129}},\ \bibinfo {pages}
  {367--438} (\bibinfo {year} {1985})}\BibitemShut {NoStop}%
\bibitem [{\citenamefont {Fyodorov}\ and\ \citenamefont
  {Mirlin}(1994)}]{fyodorov1994statistical}%
  \BibitemOpen
  \bibfield  {author} {\bibinfo {author} {\bibfnamefont {Y.~V.}\ \bibnamefont
  {Fyodorov}}\ and\ \bibinfo {author} {\bibfnamefont {A.~D.}\ \bibnamefont
  {Mirlin}},\ }\bibfield  {title} {\enquote {\bibinfo {title} {Statistical
  properties of eigenfunctions of random quasi 1{D} one-particle
  {H}amiltonians},}\ }\href {\doibase 10.1142/S0217979294001640} {\bibfield
  {journal} {\bibinfo  {journal} {Int. J. Mod. Phys. B}\ }\textbf {\bibinfo
  {volume} {8}},\ \bibinfo {pages} {3795--3842} (\bibinfo {year}
  {1994})}\BibitemShut {NoStop}%
\bibitem [{\citenamefont {Feller}(2008)}]{feller2008introduction}%
  \BibitemOpen
  \bibfield  {author} {\bibinfo {author} {\bibfnamefont {W.}~\bibnamefont
  {Feller}},\ }\href@noop {} {\emph {\bibinfo {title} {An introduction to
  Probability Theory and its Applications}}},\ \bibinfo {edition} {2nd}\ ed.,\
  Vol.~\bibinfo {volume} {2}\ (\bibinfo  {publisher} {John Wiley \& Sons},\
  \bibinfo {year} {2008})\BibitemShut {NoStop}%
\bibitem [{\citenamefont {Eisenhart}(1934)}]{eisenhart1934separable}%
  \BibitemOpen
  \bibfield  {author} {\bibinfo {author} {\bibfnamefont {L.~P.}\ \bibnamefont
  {Eisenhart}},\ }\bibfield  {title} {\enquote {\bibinfo {title} {Separable
  systems in {E}uclidean 3-space},}\ }\href {\doibase 10.1103/PhysRev.45.427.2}
  {\bibfield  {journal} {\bibinfo  {journal} {Phys. Rev.}\ }\textbf {\bibinfo
  {volume} {45}},\ \bibinfo {pages} {427} (\bibinfo {year} {1934})}\BibitemShut
  {NoStop}%
\bibitem [{\citenamefont {Eisenhart}(1948)}]{eisenhart1948enumeration}%
  \BibitemOpen
  \bibfield  {author} {\bibinfo {author} {\bibfnamefont {L.~P.}\ \bibnamefont
  {Eisenhart}},\ }\bibfield  {title} {\enquote {\bibinfo {title} {Enumeration
  of potentials for which one-particle {S}chr{\"o}dinger equations are
  separable},}\ }\href {\doibase 10.1103/PhysRev.74.87} {\bibfield  {journal}
  {\bibinfo  {journal} {Phys. Rev.}\ }\textbf {\bibinfo {volume} {74}},\
  \bibinfo {pages} {87} (\bibinfo {year} {1948})}\BibitemShut {NoStop}%
\bibitem [{\citenamefont {Olver}(1965)}]{olver1965bessel}%
  \BibitemOpen
  \bibfield  {author} {\bibinfo {author} {\bibfnamefont {F.~W.~J.}\
  \bibnamefont {Olver}},\ }\bibfield  {title} {\enquote {\bibinfo {title}
  {Bessel functions of integer order},}\ }in\ \href@noop {} {\emph {\bibinfo
  {booktitle} {Handbook of Mathematical Functions}}},\ \bibinfo {editor}
  {edited by\ \bibinfo {editor} {\bibfnamefont {M.}~\bibnamefont {Abramowitz}}\
  and\ \bibinfo {editor} {\bibfnamefont {I.~A.}\ \bibnamefont {Stegun}}}\
  (\bibinfo  {publisher} {Dover Publications, New York},\ \bibinfo {year}
  {1965})\ Chap.~\bibinfo {chapter} {9}\BibitemShut {NoStop}%
\bibitem [{\citenamefont {Paris}(2010)}]{olver2010nist}%
  \BibitemOpen
  \bibfield  {author} {\bibinfo {author} {\bibfnamefont {R.~B.}\ \bibnamefont
  {Paris}},\ }\bibfield  {title} {\enquote {\bibinfo {title} {Incomplete gamma
  and related functions},}\ }in\ \href@noop {} {\emph {\bibinfo {booktitle}
  {NIST Handbook of Mathematical Functions}}},\ \bibinfo {editor} {edited by\
  \bibinfo {editor} {\bibfnamefont {F.~W.}\ \bibnamefont {Olver}}, \bibinfo
  {editor} {\bibfnamefont {D.~W.}\ \bibnamefont {Lozier}}, \bibinfo {editor}
  {\bibfnamefont {R.~F.}\ \bibnamefont {Boisvert}}, \ and\ \bibinfo {editor}
  {\bibfnamefont {C.~W.}\ \bibnamefont {Clark}}}\ (\bibinfo  {publisher}
  {Cambridge University Press},\ \bibinfo {year} {2010})\ Chap.~\bibinfo
  {chapter} {8}, p.\ \bibinfo {pages} {173}\BibitemShut {NoStop}%
\bibitem [{\citenamefont {Gradshteyn}\ and\ \citenamefont
  {Ryzhik}(2014)}]{gradshteyn2014table}%
  \BibitemOpen
  \bibfield  {author} {\bibinfo {author} {\bibfnamefont {I.~S.}\ \bibnamefont
  {Gradshteyn}}\ and\ \bibinfo {author} {\bibfnamefont {I.~M.}\ \bibnamefont
  {Ryzhik}},\ }\href {\doibase 10.1016/c2010-0-64839-5} {\emph {\bibinfo
  {title} {Table of integrals, series, and products}}},\ edited by\ \bibinfo
  {editor} {\bibfnamefont {A.}~\bibnamefont {Jeffrey}}\ and\ \bibinfo {editor}
  {\bibfnamefont {D.}~\bibnamefont {Zwillinger}}\ (\bibinfo  {publisher}
  {Academic Press},\ \bibinfo {year} {2014})\BibitemShut {NoStop}%
\bibitem [{\citenamefont {Hammersley}(1953)}]{hammersley1953tables}%
  \BibitemOpen
  \bibfield  {author} {\bibinfo {author} {\bibfnamefont {J.~M.}\ \bibnamefont
  {Hammersley}},\ }\bibfield  {title} {\enquote {\bibinfo {title} {Tables of
  complete elliptic integrals},}\ }\href {\doibase 10.6028/jres.050.007}
  {\bibfield  {journal} {\bibinfo  {journal} {J. Res. Natl. Bur. Stand.}\
  }\textbf {\bibinfo {volume} {50}},\ \bibinfo {pages} {43} (\bibinfo {year}
  {1953})}\BibitemShut {NoStop}%
\bibitem [{\citenamefont {Beugeling}\ \emph {et~al.}(2017)\citenamefont
  {Beugeling}, \citenamefont {B{\"a}cker}, \citenamefont {Moessner},\ and\
  \citenamefont {Haque}}]{beugeling2017statistical}%
  \BibitemOpen
  \bibfield  {author} {\bibinfo {author} {\bibfnamefont {W.}~\bibnamefont
  {Beugeling}}, \bibinfo {author} {\bibfnamefont {A.}~\bibnamefont
  {B{\"a}cker}}, \bibinfo {author} {\bibfnamefont {R.}~\bibnamefont
  {Moessner}}, \ and\ \bibinfo {author} {\bibfnamefont {M.}~\bibnamefont
  {Haque}},\ }\bibfield  {title} {\enquote {\bibinfo {title} {Statistical
  properties of eigenstate amplitudes in complex quantum systems},}\ }\href
  {https://arxiv.org/abs/1710.11433v1} {\  (\bibinfo {year} {2017})},\ \bibinfo
  {note} {{a}rXiv preprint 1710.11433}\BibitemShut {NoStop}%
\bibitem [{\citenamefont {Jain}\ and\ \citenamefont
  {Samajdar}(2017)}]{jain2017nodal}%
  \BibitemOpen
  \bibfield  {author} {\bibinfo {author} {\bibfnamefont {S.~R.}\ \bibnamefont
  {Jain}}\ and\ \bibinfo {author} {\bibfnamefont {R.}~\bibnamefont
  {Samajdar}},\ }\bibfield  {title} {\enquote {\bibinfo {title} {Nodal
  portraits of quantum billiards: Domains, lines, and statistics},}\ }\href
  {\doibase 10.1103/RevModPhys.89.045005} {\bibfield  {journal} {\bibinfo
  {journal} {Rev. Mod. Phys.}\ }\textbf {\bibinfo {volume} {89}},\ \bibinfo
  {pages} {045005} (\bibinfo {year} {2017})}\BibitemShut {NoStop}%
\bibitem [{\citenamefont {Schachner}\ and\ \citenamefont
  {Obermair}(1994)}]{Schachner1994}%
  \BibitemOpen
  \bibfield  {author} {\bibinfo {author} {\bibfnamefont {H.~C.}\ \bibnamefont
  {Schachner}}\ and\ \bibinfo {author} {\bibfnamefont {G.~M.}\ \bibnamefont
  {Obermair}},\ }\bibfield  {title} {\enquote {\bibinfo {title} {Quantum
  billiards in the shape of right triangles},}\ }\href {\doibase
  10.1007/BF01316851} {\bibfield  {journal} {\bibinfo  {journal} {Z. Phys. B}\
  }\textbf {\bibinfo {volume} {95}},\ \bibinfo {pages} {113--119} (\bibinfo
  {year} {1994})}\BibitemShut {NoStop}%
\bibitem [{\citenamefont {Kaufman}, \citenamefont {Kosztin},\ and\
  \citenamefont {Schulten}(1999)}]{kaufman1999}%
  \BibitemOpen
  \bibfield  {author} {\bibinfo {author} {\bibfnamefont {D.~L.}\ \bibnamefont
  {Kaufman}}, \bibinfo {author} {\bibfnamefont {I.}~\bibnamefont {Kosztin}}, \
  and\ \bibinfo {author} {\bibfnamefont {K.}~\bibnamefont {Schulten}},\
  }\bibfield  {title} {\enquote {\bibinfo {title} {Expansion method for
  stationary states of quantum billiards},}\ }\href {\doibase 10.1119/1.19208}
  {\bibfield  {journal} {\bibinfo  {journal} {Am. J. Phys.}\ }\textbf {\bibinfo
  {volume} {67}},\ \bibinfo {pages} {133--141} (\bibinfo {year}
  {1999})}\BibitemShut {NoStop}%
\bibitem [{\citenamefont {Uhlenbeck}(1976)}]{uhlenbeck1976generic}%
  \BibitemOpen
  \bibfield  {author} {\bibinfo {author} {\bibfnamefont {K.}~\bibnamefont
  {Uhlenbeck}},\ }\bibfield  {title} {\enquote {\bibinfo {title} {Generic
  properties of eigenfunctions},}\ }\href {\doibase 10.2307/2374041} {\bibfield
   {journal} {\bibinfo  {journal} {Am. J. Math.}\ }\textbf {\bibinfo {volume}
  {98}},\ \bibinfo {pages} {1059--1078} (\bibinfo {year} {1976})}\BibitemShut
  {NoStop}%
\bibitem [{\citenamefont {Aronovitch}\ \emph {et~al.}(2012)\citenamefont
  {Aronovitch}, \citenamefont {Band}, \citenamefont {Fajman},\ and\
  \citenamefont {Gnutzmann}}]{aronovitch2012nodal}%
  \BibitemOpen
  \bibfield  {author} {\bibinfo {author} {\bibfnamefont {A.}~\bibnamefont
  {Aronovitch}}, \bibinfo {author} {\bibfnamefont {R.}~\bibnamefont {Band}},
  \bibinfo {author} {\bibfnamefont {D.}~\bibnamefont {Fajman}}, \ and\ \bibinfo
  {author} {\bibfnamefont {S.}~\bibnamefont {Gnutzmann}},\ }\bibfield  {title}
  {\enquote {\bibinfo {title} {Nodal domains of a non-separable problem---the
  right-angled isosceles triangle},}\ }\href {\doibase
  10.1088/1751-8113/45/8/085209} {\bibfield  {journal} {\bibinfo  {journal} {J.
  Phys. A: Math. Theor.}\ }\textbf {\bibinfo {volume} {45}},\ \bibinfo {pages}
  {085209} (\bibinfo {year} {2012})}\BibitemShut {NoStop}%
\bibitem [{\citenamefont {Brack}\ and\ \citenamefont
  {Bhaduri}(2003)}]{brack2003semiclassical}%
  \BibitemOpen
  \bibfield  {author} {\bibinfo {author} {\bibfnamefont {M.}~\bibnamefont
  {Brack}}\ and\ \bibinfo {author} {\bibfnamefont {R.~K.}\ \bibnamefont
  {Bhaduri}},\ }\href@noop {} {\emph {\bibinfo {title} {Semiclassical
  physics}}},\ Vol.~\bibinfo {volume} {96}\ (\bibinfo  {publisher} {Westview
  Press},\ \bibinfo {address} {Boulder, CO},\ \bibinfo {year}
  {2003})\BibitemShut {NoStop}%
\bibitem [{\citenamefont {McCartin}(2003)}]{mccartin2003eigenstructure}%
  \BibitemOpen
  \bibfield  {author} {\bibinfo {author} {\bibfnamefont {B.~J.}\ \bibnamefont
  {McCartin}},\ }\bibfield  {title} {\enquote {\bibinfo {title} {Eigenstructure
  of the equilateral triangle, {P}art {I}: {T}he {D}irichlet problem},}\ }\href
  {\doibase 10.1137/S003614450238720} {\bibfield  {journal} {\bibinfo
  {journal} {SIAM Rev.}\ }\textbf {\bibinfo {volume} {45}},\ \bibinfo {pages}
  {267--287} (\bibinfo {year} {2003})}\BibitemShut {NoStop}%
\bibitem [{\citenamefont {Jain}, \citenamefont {Gr\'emaud},\ and\ \citenamefont
  {Khare}(2002)}]{jain2002}%
  \BibitemOpen
  \bibfield  {author} {\bibinfo {author} {\bibfnamefont {S.~R.}\ \bibnamefont
  {Jain}}, \bibinfo {author} {\bibfnamefont {B.}~\bibnamefont {Gr\'emaud}}, \
  and\ \bibinfo {author} {\bibfnamefont {A.}~\bibnamefont {Khare}},\ }\bibfield
   {title} {\enquote {\bibinfo {title} {Quantum modes on chaotic motion:
  Analytically exact results},}\ }\href {\doibase 10.1103/PhysRevE.66.016216}
  {\bibfield  {journal} {\bibinfo  {journal} {Phys. Rev. E}\ }\textbf {\bibinfo
  {volume} {66}},\ \bibinfo {pages} {016216} (\bibinfo {year}
  {2002})}\BibitemShut {NoStop}%
\bibitem [{\citenamefont {Samajdar}\ and\ \citenamefont
  {Jain}(2014{\natexlab{a}})}]{samajdar2014JPA}%
  \BibitemOpen
  \bibfield  {author} {\bibinfo {author} {\bibfnamefont {R.}~\bibnamefont
  {Samajdar}}\ and\ \bibinfo {author} {\bibfnamefont {S.~R.}\ \bibnamefont
  {Jain}},\ }\bibfield  {title} {\enquote {\bibinfo {title} {Nodal domains of
  the equilateral triangle billiard},}\ }\href {\doibase
  10.1088/1751-8113/47/19/195101} {\bibfield  {journal} {\bibinfo  {journal}
  {J. Phys. A: Math. Theor.}\ }\textbf {\bibinfo {volume} {47}},\ \bibinfo
  {pages} {195101} (\bibinfo {year} {2014}{\natexlab{a}})}\BibitemShut
  {NoStop}%
\bibitem [{\citenamefont {Samajdar}\ and\ \citenamefont
  {Jain}(2014{\natexlab{b}})}]{samajdar2014nodal}%
  \BibitemOpen
  \bibfield  {author} {\bibinfo {author} {\bibfnamefont {R.}~\bibnamefont
  {Samajdar}}\ and\ \bibinfo {author} {\bibfnamefont {S.~R.}\ \bibnamefont
  {Jain}},\ }\bibfield  {title} {\enquote {\bibinfo {title} {A nodal domain
  theorem for integrable billiards in two dimensions},}\ }\href {\doibase
  10.1016/j.aop.2014.08.010} {\bibfield  {journal} {\bibinfo  {journal} {Ann.
  Phys.}\ }\textbf {\bibinfo {volume} {351}},\ \bibinfo {pages} {1--12}
  (\bibinfo {year} {2014}{\natexlab{b}})}\BibitemShut {NoStop}%
\bibitem [{\citenamefont {Manjunath}, \citenamefont {Samajdar},\ and\
  \citenamefont {Jain}(2016)}]{manjunath2016difference}%
  \BibitemOpen
  \bibfield  {author} {\bibinfo {author} {\bibfnamefont {N.}~\bibnamefont
  {Manjunath}}, \bibinfo {author} {\bibfnamefont {R.}~\bibnamefont {Samajdar}},
  \ and\ \bibinfo {author} {\bibfnamefont {S.~R.}\ \bibnamefont {Jain}},\
  }\bibfield  {title} {\enquote {\bibinfo {title} {A difference-equation
  formalism for the nodal domains of separable billiards},}\ }\href {\doibase
  10.1016/j.aop.2016.04.014} {\bibfield  {journal} {\bibinfo  {journal} {Ann.
  Phys.}\ }\textbf {\bibinfo {volume} {372}},\ \bibinfo {pages} {68--73}
  (\bibinfo {year} {2016})}\BibitemShut {NoStop}%
\bibitem [{\citenamefont {Mandwal}\ and\ \citenamefont
  {Jain}(2017)}]{mandwal2017}%
  \BibitemOpen
  \bibfield  {author} {\bibinfo {author} {\bibfnamefont {A.~K.}\ \bibnamefont
  {Mandwal}}\ and\ \bibinfo {author} {\bibfnamefont {S.~R.}\ \bibnamefont
  {Jain}},\ }\bibfield  {title} {\enquote {\bibinfo {title} {Raising and
  lowering operators for quantum billiards},}\ }\href {\doibase
  10.1007/s12043-017-1432-x} {\bibfield  {journal} {\bibinfo  {journal}
  {Pramana - J. Phys.}\ }\textbf {\bibinfo {volume} {89}},\ \bibinfo {pages}
  {35} (\bibinfo {year} {2017})}\BibitemShut {NoStop}%
\bibitem [{\citenamefont {Robinett}(1996)}]{Robinett1}%
  \BibitemOpen
  \bibfield  {author} {\bibinfo {author} {\bibfnamefont {R.~W.}\ \bibnamefont
  {Robinett}},\ }\bibfield  {title} {\enquote {\bibinfo {title} {Visualizing
  the solutions for the circular infinite well in quantum and classical
  mechanics},}\ }\href {\doibase 10.1119/1.18188} {\bibfield  {journal}
  {\bibinfo  {journal} {Am. J. Phys.}\ }\textbf {\bibinfo {volume} {64}},\
  \bibinfo {pages} {440--446} (\bibinfo {year} {1996})}\BibitemShut {NoStop}%
\bibitem [{\citenamefont {Robinett}(2003)}]{0143-0807-24-3-302}%
  \BibitemOpen
  \bibfield  {author} {\bibinfo {author} {\bibfnamefont {R.~W.}\ \bibnamefont
  {Robinett}},\ }\bibfield  {title} {\enquote {\bibinfo {title} {Quantum
  mechanics of the two-dimensional circular billiard plus baffle system and
  half-integral angular momentum},}\ }\href {\doibase
  10.1088/0143-0807/24/3/302} {\bibfield  {journal} {\bibinfo  {journal} {Eur.
  J. Phys.}\ }\textbf {\bibinfo {volume} {24}},\ \bibinfo {pages} {231}
  (\bibinfo {year} {2003})}\BibitemShut {NoStop}%
\bibitem [{\citenamefont {Olver}(1952)}]{olver1952some}%
  \BibitemOpen
  \bibfield  {author} {\bibinfo {author} {\bibfnamefont {F.~W.~J.}\
  \bibnamefont {Olver}},\ }\bibfield  {title} {\enquote {\bibinfo {title} {Some
  new asymptotic expansions for {B}essel functions of large orders},}\ }in\
  \href {\doibase 10.1017/S030500410002781X} {\emph {\bibinfo {booktitle}
  {Math. Proc. Cambridge Phil. Soc.}}},\ Vol.~\bibinfo {volume} {48}\ (\bibinfo
  {year} {1952})\ pp.\ \bibinfo {pages} {414--427}\BibitemShut {NoStop}%
\end{thebibliography}%

\end{document}